\documentclass[aps,prl,twocolumn,superscriptaddress,longbibliography,floatfix]{revtex4-2}
\usepackage{amsmath,amssymb,graphicx,braket,dsfont,bm}
\usepackage[colorlinks=true,allcolors=blue]{hyperref}
\usepackage{xcolor}
\usepackage{xeCJK}
\usepackage[T1]{fontenc}
\usepackage[utf8]{inputenc}
\newcommand{\ii}{\text{i}}
\begin{document}

\title{Universal Free-Electron-Sideband Control} \author{Tong Shen (沈彤)} \affiliation{School of Physics, Wuhan University, Wuhan 430072, China} \author{Zhexin Zhao} \email{zhexin.zhao@fau.de} \affiliation{Department of Physics, Friedrich-Alexander University (FAU) Erlangen--N\"urnberg, Staudtstra\ss e 1, 91058 Erlangen, Germany} \author{Meng Xiao} \email{phmxiao@whu.edu.cn} \affiliation{School of Physics, Wuhan University, Wuhan 430072, China} \affiliation{Wuhan Institute of Quantum Technology, Wuhan 430206, China}

\begin{abstract}
Advances in free-electron quantum optics are establishing free electrons as a versatile quantum resource. Photon-induced near-field electron microscopy (PINEM) interactions and free drift provide elementary operations for shaping the longitudinal electron wavefunction, yet their full control capabilities remain unexplored. Here, we prove that a finite sequence of alternating monochromatic PINEM interactions and free drifts can approximate any prescribed inner-product-preserving map from a finite-dimensional input subspace into the full ideal electron-sideband Hilbert space to arbitrary accuracy. We therefore call the protocol PINEMAD (PINEM And Drift). This framework encompasses state preparation, unitary gates, and isometric encoding. We further show numerically that parameter optimization enables high-fidelity state preparation at finite electron energy spreads, using experimentally accessible coupling strengths and tabletop-scale total drift lengths. These results establish electron energy sidebands as a programmable quantum resource.
\end{abstract}

\maketitle

\textit{Introduction.}---The emerging field of free-electron quantum optics (FEQO)~\cite{ruimyFreeelectronQuantumOptics2025,garciadeabajoRoadmapQuantumNanophotonics2025} has expanded the role of flying electrons from probes of nanoscale excitations~\cite{garciadeabajoProbingPhotonicLocal2008,polmanElectronbeamSpectroscopyNanophotonics2019a,arquelopezAtomicFloquetPhysics2022} to controllable quantum resources carrying quantum information~\cite{kfirEntanglementsElectronsCavity2019b,dahanImprintingQuantumStatistics2021,feistCavitymediatedElectronphotonPairs2022,arendElectronsHeraldNonclassical2025,pescollerCouplingFreeElectrons2026}. Beyond quantum-information encoding, electron wavefunction-shaping~\cite{vanacoreAttosecondCoherentControl2018,reinhardtTheoryShapingElectron2020a,giulioFreeelectronShapingUsing2020,garciadeabajoOpticalModulationElectron2021} has enabled a range of experimental and theoretical advances, including the generation of attosecond electron pulse trains~\cite{priebeAttosecondElectronPulse2017,morimotoDiffractionMicroscopyAttosecond2018,yaluninTailoredHighcontrastAttosecond2021}, probing quantum coherence in matter~\cite{ruimyAtomicResolutionQuantumMeasurements2021,zhangQuantumStateInterrogation2022,karnieliQuantumSensingStrongly2023}, controlling atomic and material excitations~\cite{goverFreeElectronBoundElectronResonant2020,zhaoQuantumEntanglementModulation2021,konecnaEntanglingFreeElectrons2022,abad-arredondoQuantumStatePreparation2025,wuCoherentControlThreeLevel2026,shenHeraldedFreeElectronWriting2026}, engineering photonic quantum states~\cite{kfirOpticalCoherenceTransfer2021,baranesFreeElectronsInduce2022,huangElectronPhotonQuantumState2023,dahanCreationOpticalCat2023,baranesFreeelectronInteractionsPhotonic2023,giulioTunableQuantumLight2025}, and controlling radiation processes~\cite{wongControlQuantumElectrodynamical2021,limQuantumInterferenceFundamentally2023,wongFreeelectronCrystalsEnhanced2024}. For the longitudinal degree of freedom, two elementary operations for shaping the electron wavefunction are photon-induced near-field electron microscopy (PINEM) interactions~\cite{barwickPhotoninducedNearfieldElectron2009,garciadeabajoMultiphotonAbsorptionEmission2010a,parkPhotoninducedNearfieldElectron2010,parkRelativisticEffectsPhotonInduced2012,feistQuantumCoherentOptical2015,echternkampRamseytypePhaseControl2016,shilohQuantumCoherentLightElectronInteraction2022} and free drift~\cite{baumQuantumDynamicsAttosecond2017,yaluninTailoredHighcontrastAttosecond2021,tsarevFreeelectronQubitsMaximumcontrast2021}. In PINEM, coherent interaction with a monochromatic optical near field produces electron energy sidebands separated by the photon energy, while free drift changes the relative phases between these sidebands. Beyond sideband-state shaping~\cite{giulioModulationCathodoluminescence2021,morimotoCoherentScatteringOptically2021,garciadeabajoSpatiotemporalElectronBeam2023,zhaoElectronPulseCompression2021,yaluninTailoredHighcontrastAttosecond2021,priebeAttosecondElectronPulse2017}, these operations have also been used to construct quantum gates for electron-sideband states~\cite{reinhardtFreeElectronQubits2021,tsarevFreeelectronQubitsMaximumcontrast2021,braimanSyntheticHilbertSpace2023}. These advances raise a fundamental question: are monochromatic PINEM interactions and free drift alone sufficient for universal coherent control of electron energy sidebands?

In this {work}, we establish that a finite sequence of alternating monochromatic PINEM interactions and free drifts can approximate any prescribed inner-product-preserving map from an arbitrary finite-dimensional input subspace into the full ideal sideband Hilbert space to arbitrary accuracy. We call this control protocol PINEMAD (PINEM And Drift). This result establishes its universality for state preparation, unitary gates, and isometric encoding of electron sidebands. Numerical optimizations demonstrate near-unity fidelities for representative examples of all three using short PINEMAD sequences. We further demonstrate numerically that PINEMAD remains effective for state preparation at realistic finite electron energy spreads, and examine the coupling strengths and drift lengths required for its implementation.

\begin{figure}[t]
    \centering
    \includegraphics[width=\columnwidth]{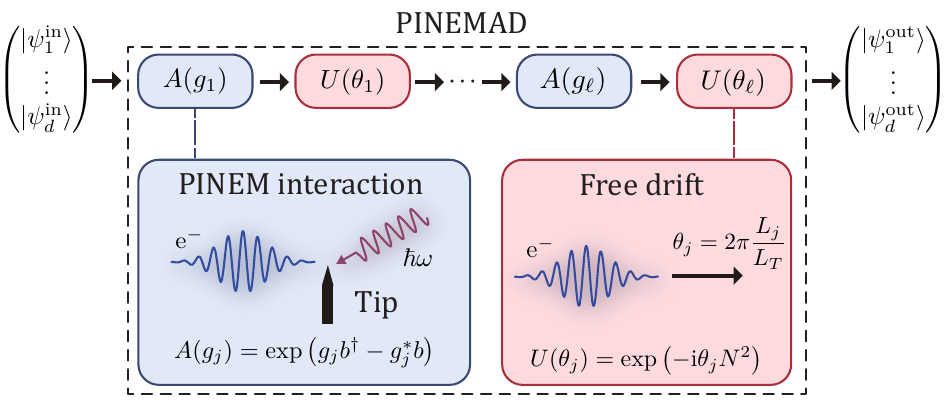}
\caption{Schematic of an $\ell$-stage PINEMAD sequence. A single sequence of alternating PINEM interactions $A(g_j)$ and free drifts $U(\theta_j)$ acts on a set of $d$ input states $\{\ket{\psi_j^{\mathrm{in}}}\}_{j=1}^{d}$ and approximately realizes the prescribed mapping $\{\ket{\psi_j^{\mathrm{in}}}\}_{j=1}^{d}\rightarrow\{\ket{\psi_j^{\mathrm{out}}}\}_{j=1}^{d}$. Each PINEM interaction is controlled by the complex coupling $g_j$, while a free-drift section of length $L_j$ produces the phase $\theta_j=2\pi L_j/L_T$, with $L_T$ the Talbot length.}
    \label{fig:1}
\end{figure}

\textit{Universality of PINEMAD.}---When only the longitudinal dynamics are considered, an initially monochromatic free electron with reference kinetic energy $E_0$, undergoing PINEM interactions with a monochromatic light field of angular frequency $\omega_0$ and free drift along the $z$ direction, is described in the energy-sideband basis $\ket{n}\equiv\ket{E_0+n\hbar\omega_0}$.  The corresponding Hilbert space spanned by these basis states is $\mathcal H=\ell^2(\mathbb Z)$. In this basis, the PINEM interaction and free drift are described by the operators
\begin{equation}\label{operators}
    A(g)=\exp\left(gb^{\dagger}-g^*b\right),\;\; U(\theta)=\exp\left(-\mathrm{i}\theta N^2\right),
\end{equation}
where $b$ is the sideband lowering operator, $b\ket{n}=\ket{n-1}$, and $N$ is the sideband-index operator, $N\ket{n}=n\ket{n}$. The complex coupling $g$ characterizes the PINEM interaction, while the dimensionless drift phase $\theta=2\pi L/L_T$ is determined by the free-drift length $L$ and the electron Talbot length $L_T$~\cite{baumQuantumDynamicsAttosecond2017,yaluninTailoredHighcontrastAttosecond2021,tsarevFreeelectronQubitsMaximumcontrast2021}. The derivation and validity conditions of these operators are provided in Sec.~I of the {Supplementary Information}. Alternating $\ell$ PINEM interactions with $\ell$ free drifts gives the $\ell$-stage PINEMAD sequence shown in Fig.~\hyperref[fig:1]{1}:
\begin{equation}
    W_\ell:=U(\theta_\ell)A(g_\ell)\cdots U(\theta_1)A(g_1).
\end{equation}
where the complex couplings $g_j$ and drift phases $\theta_j$, $j=1,\ldots,\ell$, constitute the control parameters. We then establish the following theorem.

\textit{Theorem 1.}---Let $\mathcal E\subset \mathcal H$ be any finite-dimensional subspace. Then, for any linear inner-product-preserving operator $T:\mathcal E\rightarrow \mathcal H$, and any $\varepsilon>0$, there exist a finite integer $\ell$ and parameters $\{g_j,\theta_j\}_{j=1}^{\ell}$ such that
\begin{equation}
    \left\|(W_\ell-T)P_{\mathcal E}\right\|<\varepsilon.
\end{equation}
where $\|\cdot\|$ denotes the operator norm and $P_{\mathcal E}$ is the projector onto $\mathcal E$. 

We first clarify the meanings of $\mathcal E$ and $T$. The input subspace $\mathcal E$ is spanned by $d$ ($d\equiv \dim\,\mathcal E$) linearly independent states $\{\ket{\psi_j}\}_{j=1}^{d}\subset\mathcal H$. These states need not be mutually orthogonal, and each $\ket{\psi_j}$ may have finitely or infinitely many populated sidebands, provided it is normalizable. The linear map $T:\mathcal E\to\mathcal H$ preserves inner products, meaning that $\braket{T\psi|T\phi}=\braket{\psi|\phi}$ for all $\ket{\psi},\ket{\phi}\in\mathcal E$. Its image $T(\mathcal E)$ can be any $d$-dimensional subspace of $\mathcal H$, either identical to or different from $\mathcal E$.

\begin{figure}[t]
    \centering
    \includegraphics[width=\columnwidth]{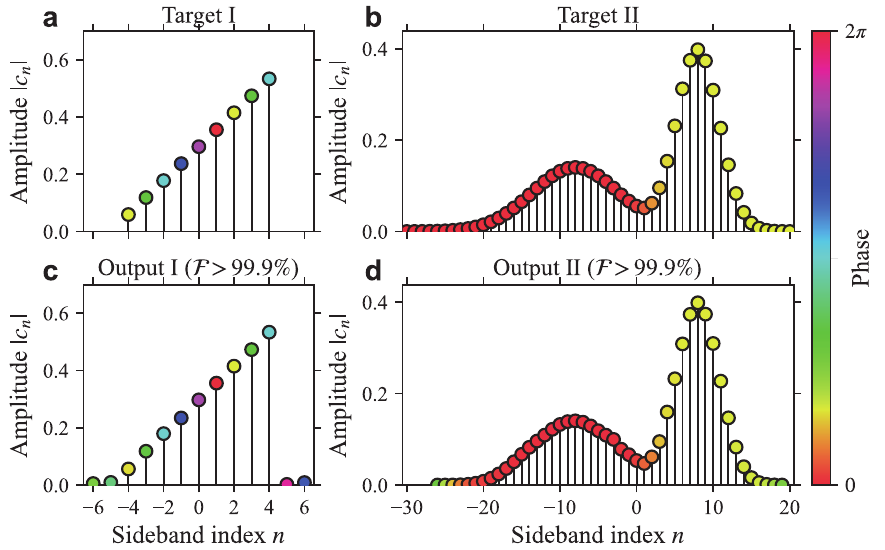}
\caption{Arbitrary electron-sideband state synthesis using PINEMAD. (a) Target I, nine sidebands $n=-4,\ldots,4$ with linearly increasing amplitudes and a neighboring phase increment of $\pi/3$. (b) Target II, a coherent superposition of two Gaussian-type sideband wavepackets. (c),(d) Corresponding output states synthesized from the initial state $\ket{0}$ using $\ell=6$ and $\ell =8$ PINEMAD sequences, with fidelities $\mathcal{F}>99.9\%$. The stem heights represent the amplitudes $|c_n|$, while the colors indicate the phases of the coefficients $c_n$.}
    \label{fig:2}
\end{figure}

\begin{figure}[t]
    \centering
    \includegraphics[width=\columnwidth]{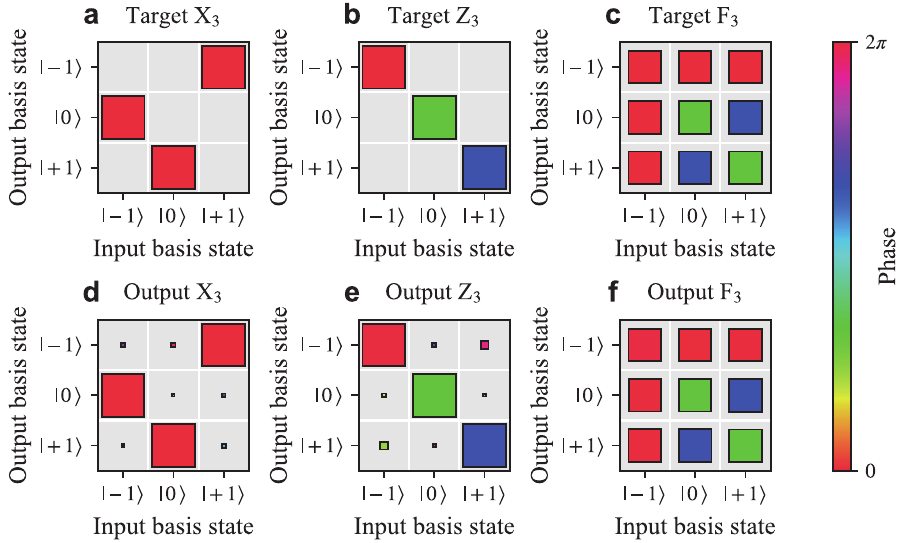}
\caption{Qutrit-gate synthesis with PINEMAD. (a)--(c) Target matrices of $X_3$, $Z_3$, and $F_3$ on $\mathcal E=\operatorname{span}\{\ket{-1},\ket{0},\ket{+1}\}$. (d)--(f) Corresponding matrices of $P_{\mathcal E}W_\ell P_{\mathcal E}$. Columns and rows label input and output basis states, respectively. Square area is proportional to the matrix-element magnitude, while color indicates its phase with the global phase removed. This area scaling makes small residual elements visually apparent even at gate fidelities above $99\%$.}
    \label{fig:3}
\end{figure}

\begin{figure}[t]
    \centering
    \includegraphics[width=\columnwidth]{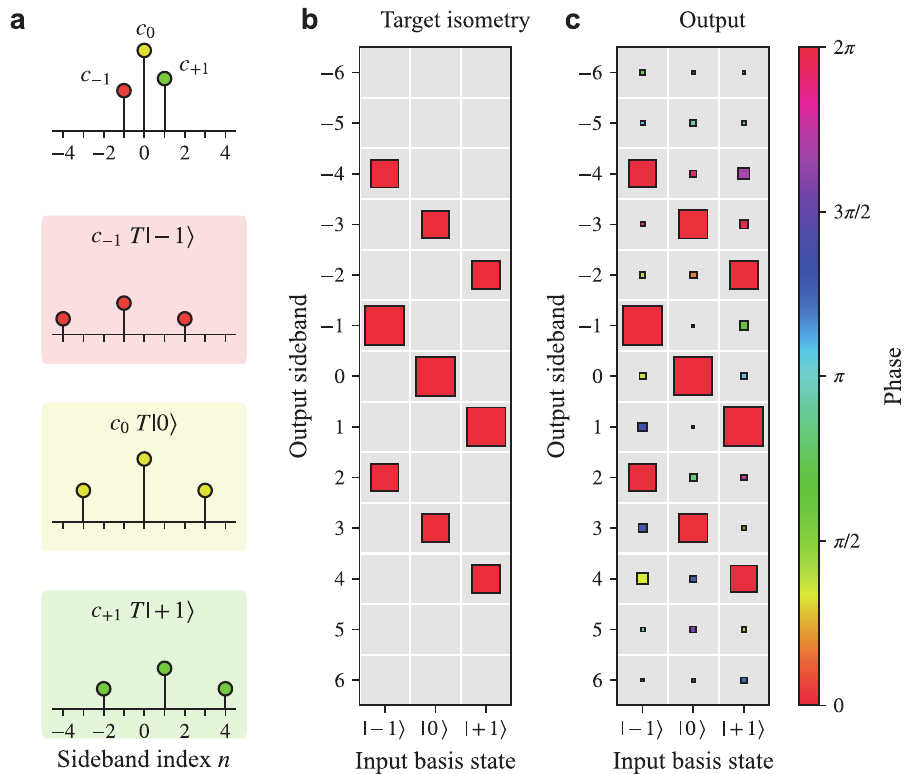}
\caption{Isometric encoding of a sideband qutrit. (a) Schematic of the isometry defined in Eq.~(5). (b) Matrix elements $\braket{n|T|m}$ of the target isometry. (c) Corresponding matrix elements $\bra{n}W_\ell\ket{m}$. Columns label input basis states and rows label output sidebands. Square areas and colors follow the conventions of Fig.~3. The isometry fidelity exceeds $99\%$.}
    \label{fig:4}
\end{figure}

For $d=1$, the theorem enables us to prepare any state in $\mathcal H$ from an arbitrary initial state in $\mathcal H$ (see Fig.~\hyperref[fig:2]{2}), which is one of the central goals in FEQO. For $d>1$ and $T(\mathcal E)=\mathcal E$, PINEMAD can approximate an arbitrary unitary transformation on $\mathcal E$ and can therefore be used, e.g., to construct quantum gates for free-electron sideband states (see Fig.~\hyperref[fig:3]{3}). More generally, without requiring $T(\mathcal E)=\mathcal E$, PINEMAD can perform an isometric encoding, i.e., transform an arbitrary state $\sum_{j=1}^{d}\alpha_j\ket{\psi_j}$ into $\sum_{j=1}^{d}\alpha_j\ket{\phi_j}$, where $\ket{\phi_j}=T\ket{\psi_j}$, and the coefficients $\{\alpha_j\}$ are arbitrary and need not be {known} when designing the PINEMAD sequence (see Fig.~\hyperref[fig:4]{4}).

The origin of this control capability can be understood by comparing sequences with and without free drift. Without drift, successive PINEM interactions satisfy $\prod_{j=1}^{\ell}A(g_j)=A(\sum_{j=1}^{\ell}g_j)$ and therefore reduce to a single interaction. Interleaving PINEM interactions with free drifts generally prevents this reduction because $[U(\theta),A(g)]\neq0$. An $\ell$-stage sequence thus provides $3\ell$ tunable real control parameters: the two quadratures of each complex coupling $g_j$ and each drift phase $\theta_j$.

The proof of Theorem~1 is given in detail in Sec.~II of the {Supplementary Information} and outlined as follows. We first introduce an auxiliary continuous-time PINEMAD system and assume it satisfies the hypotheses of Theorem~2.6 of Ref.~\cite{boscainMultiinputSchrodingerEquation2014}. Applying this controllability result and using the Lie--Trotter product formula~\cite{trotterProductSemigroups1959} to approximate the resulting continuous evolution by a finite PINEMAD sequence establishes Theorem~1 under these hypotheses. We then verify that all these hypotheses indeed hold for the auxiliary PINEMAD system.

However, the theorem only guarantees the existence of such a finite sequence and does not provide a constructive procedure for finding it. In practice, for a given target operator $T$, we numerically search for the $3\ell$ control parameters of $W_\ell$. The {optimization methods, optimized PINEMAD parameters, and fidelity data} for {Figs.~2--5} are provided in Sec.~VI of the {Supplementary Information}.  {The} globally optimal approximation error is nonincreasing with $\ell$, since an additional stage can be made trivial by setting $g_{\ell+1}=\theta_{\ell+1}=0$. By Theorem~1, this error tends to zero as $\ell\to\infty$.

\textit{Arbitrary sideband state synthesis.}---For the numerical examples below, we consider a monochromatic incident electron $\ket{0}=\ket{E_0}$. To demonstrate the capability of PINEMAD in state preparation, we consider two representative targets [see Fig.~\hyperref[fig:2]{2(a)-(b)}]. Target I is a nine-sideband state $\ket{\psi_{\mathrm{I}}}\propto\sum_{k=1}^{9}k\exp(\mathrm{i}k\pi/3)\ket{k-5}$. Target II is a normalized coherent superposition of two Gaussian-type sideband wavepackets, $\ket{\psi_{\mathrm{II}}}\propto\ket{G(-8,4)}+2\exp(\mathrm{i}\pi/3)\ket{G(8,2)}$, where $\ket{G(n_0,\sigma)}$ is the normalized state proportional to $\sum_n\exp[-(n-n_0)^2/(4\sigma^2)]\ket{n}$, with center $n_0$ and width $\sigma$. Unlike Target I, Target II has a broader sideband distribution with infinite and rapidly decaying Gaussian tails. For an $\ell$-stage PINEMAD sequence, we optimize the $3\ell$ control parameters under the constraints $|g_j|\leq2$ and $\theta_j\in[0,2\pi)$, using the state fidelity $\mathcal F:=|\bra{\psi_{\mathrm{tar}}}W_\ell\ket{0}|^2$ as the objective. Targets I and II are synthesized from $\ket{0}$ with fidelities above $99.9\%$ using $\ell=6$ and $\ell=8$ stages, respectively. Details of the optimized parameters and fidelity data for {Figs.~2--5} are provided in Sec.~VI of the {Supplementary Information}. 

Pre-shaped free electrons have been proposed as resources for the heralded preparation and shaping of nonclassical light~\cite{dahanCreationOpticalCat2023,giulioTunableQuantumLight2025,huangElectronPhotonQuantumState2023,zhuProgrammingCoherentQuantum2026}. Our result provides, in principle, a way to approximate arbitrary normalizable electron-sideband input states for exploring these electron-mediated photonic protocols. To illustrate the potential with short PINEMAD sequences, we also optimize the preparation of the two-sideband state $(\ket{0}+\exp(\ii\varphi)\ket{1})/\sqrt{2}$, with $\varphi\in[0,2\pi)$, achieving fidelities of $\mathcal F=96.30\%$ and $99.90\%$ for $\ell=2$ and $\ell=3$, respectively. These fidelities are independent of the target relative phase (see Sec.~III of the {Supplementary Information}). Such states can be used to probe the quantum state and coherence of a two-level system~\cite{ruimyAtomicResolutionQuantumMeasurements2021}. 
Unlike sideband-synthesis protocols based on time-dependent optical-phase
modulation or multiple optical frequencies
~\cite{alvarez-ahedoPhaseProgrammableFreeElectron2026},
PINEMAD establishes universal control using only PINEM interactions
at a single fixed optical frequency and free drift.

\textit{Arbitrary unitary synthesis.}---As an example, we illustrate unitary synthesis on an electron-sideband qutrit encoded in the subspace $\mathcal E=\operatorname{span}\{\ket{-1},\ket{0},\ket{+1}\}$, seeking a single PINEMAD sequence for each gate that acts on arbitrary input in $\mathcal E$. We consider the cyclic shift $X_3$, the phase gate $Z_3$, and the discrete Fourier transform $F_3$. In the ordered basis $\{\ket{-1},\ket{0},\ket{+1}\}$, their nonzero matrix elements are
\begin{equation}
    (X_3)_{j+1,j}=1,\quad (Z_3)_{jj}=\omega^j,\quad (F_3)_{jk}=\frac{\omega^{jk}}{\sqrt{3}},
\end{equation}
where $j,k=0,1,2$, the index $j+1$ in $X_3$ is understood modulo three, and $\omega=\exp(2\pi\mathrm{i}/3)$. For each target $T\in\{X_3,Z_3,F_3\}$, we optimize the coherent gate fidelity $\mathcal F_{\text{gate}}:=|\operatorname{Tr}(T^\dagger B)|^2/9$, where $B$ is the matrix of $P_{\mathcal E}W_\ell P_{\mathcal E}$ in the qutrit basis. This fidelity accounts for gate errors within $\mathcal E$ and leakage into other sidebands for inputs in $\mathcal E$. Under the constraints $|g_j|\leq 2$ and $\theta_j\in[0,2\pi)$, all three gates are synthesized with fidelities above $99\%$ using PINEMAD sequences with $\ell=7$ stages (see Fig.~\hyperref[fig:3]{3}).

\textit{Arbitrary isometry synthesis.}---Isometry synthesis enables an arbitrary electron-sideband state to be encoded into a different sideband subspace. For a state encoded in $\mathcal E=\operatorname{span}\{\ket{-1},\ket{0},\ket{+1}\}$, we target the isometry
\begin{equation}
    T\ket{m}=\frac{\ket{m-3}+2\ket{m}+\ket{m+3}}{\sqrt{6}},
\end{equation}
where $m=-1,0,+1$. The three output states are normalized and occupy disjoint sets of sidebands, making them orthonormal. We optimize the coherent isometry fidelity $\mathcal F_{\mathrm{iso}}:=|\sum_{m=-1}^{1}\bra{m}T^\dagger W_\ell\ket{m}|^2/9$. Under the constraints $|g_j|\leq1$ and $\theta_j\in[0,2\pi)$, a sequence with $\ell=10$ stages achieves $\mathcal F_{\mathrm{iso}}>99\%$ (see Fig.~\hyperref[fig:4]{4}).

Beyond information encoding, both unitary and isometric synthesis enable control over electron measurements. Unitary synthesis changes the basis of a projective energy measurement, as proposed for free-electron heralding with a PINEM interaction~\cite{huangElectronPhotonQuantumState2023}. Isometric synthesis further enables generalized measurements with more outcomes than the input-space dimension. Ideally, detecting sideband $n$ by electron energy-loss spectroscopy (EELS) after applying $T$ corresponds to the measurement effect $M_n=T^\dagger\ket{n}\bra{n}T$ on $\mathcal E$, enabling flexible design of generalized electron measurements.

\textit{Effect of finite energy spread.}---Theorem~1 applies to an ideal ladder of definite electron energies, whereas the incident electron in an experiment has a finite energy spread. We now examine its effect on state preparation. We model the incident electron as a coherent Gaussian energy wavepacket centered at the reference kinetic energy $E_0$,
\begin{equation}
    \ket{\Psi_{\mathrm{in},\sigma_E}}
    =\int_{-\infty}^{+\infty}\mathrm{d}E\,
    f(E-E_0,\sigma_E)\ket{E},
\end{equation}
where $f(\epsilon,\sigma_E) =(2\pi\sigma_E^2)^{-1/4}\exp[-\epsilon^2/(4\sigma_E^2)]$. For convenience, we introduce the continuous dimensionless energy $x=(E-E_0)/(\hbar\omega_0)$ and define $\ket{x}\equiv\ket{E_0+x\hbar\omega_0}$. The ideal sidebands correspond to $x=n$, with $n\in\mathbb Z$. For an ideal target state $\ket{\psi_{\mathrm{tar}}}=\sum_n c_n\ket{n}$, its finite-width counterpart is
\begin{equation}
    \ket{\Psi_{\mathrm{tar},\sigma_E}}=\mathcal N\int_{-\infty}^{+\infty}\mathrm{d}x\,f(x\hbar\omega_0,\sigma_E)\sum_n c_n\ket{x+n},
\end{equation}
where $\mathcal N$ normalizes the target wavepacket. We then define the finite-width state-preparation fidelity as $\mathcal F_{\sigma_E}:= |\braket{\Psi_{\mathrm{tar},\sigma_E}| \Psi_{\mathrm{out},\sigma_E}}|^2$. The derivation and expression of the fidelity are provided in Sec.~IV of the {Supplementary Information}.

To connect this analysis with experimental conditions, we note that reported incident-electron energy spreads are approximately $0.35$--$0.5\,\mathrm{eV}$~\cite{henkeIntegratedPhotonicsEnables2021,auadMeVElectronSpectromicroscopy2023}, while monochromation can reduce the spread to $30$--$50\,\mathrm{meV}$~\cite{auadMeVElectronSpectromicroscopy2023} and even $9\,\mathrm{meV}$~\cite{krivanekVibrationalSpectroscopyElectron2014}. For our Gaussian model, we characterize the energy spread by its full width at half maximum (FWHM), $\Delta E=2\sqrt{2\ln2}\,\sigma_E$. At an optical wavelength of $1030\,\mathrm{nm}$~\cite{shilohQuantumCoherentLightElectronInteraction2022}, applying the parameters optimized for the ideal sideband ladder in Fig.~\hyperref[fig:2]{2} directly to a finite-width incident wavepacket yields fidelities of only $6.30\times10^{-2}$ and $1.77\times10^{-3}$ for Targets I and II, respectively, at $\Delta E=0.1\,\mathrm{eV}$. Both fidelities remain below $10^{-2}$ at the tested widths $\Delta E=0.2$--$0.5\,\mathrm{eV}$ (see Sec.~VI of the {Supplementary Information} for the corresponding values). This loss arises primarily from energy-dependent phases accumulated during free drift, with larger drift phases generally producing greater phase variation across the incident wavepacket.

However, finite-width state preparation need not use the parameters optimized for the ideal sideband ladder. A complementary theoretical guarantee is provided by the following corollary.

\textit{Corollary 1.}---Let $\mathcal E$ and $T$ be as in Theorem~1. Denote by $\ket{\Psi_{\psi,\sigma_E}}$ and $\ket{\Psi_{T\psi,\sigma_E}}$ the normalized finite-width counterparts of $\ket{\psi}$ and $T\ket{\psi}$, respectively, constructed using their sideband coefficients and the same Gaussian envelope $f$. For any $\varepsilon>0$, there exist a finite $\ell$, PINEMAD parameters $\{g_j,\theta_j\}_{j=1}^{\ell}$, and $\sigma_E^*>0$ such that
\begin{equation}
    1-\left|\bra{\Psi_{T\psi,\sigma_E}}W_\ell\ket{\Psi_{\psi,\sigma_E}}\right|^2<\varepsilon
\end{equation}
for every normalized $\ket{\psi}\in\mathcal E$ and every $0<\sigma_E<\sigma_E^*$, with the same sequence used throughout.

\begin{figure}[h]
    \centering
    \includegraphics[width=\columnwidth]{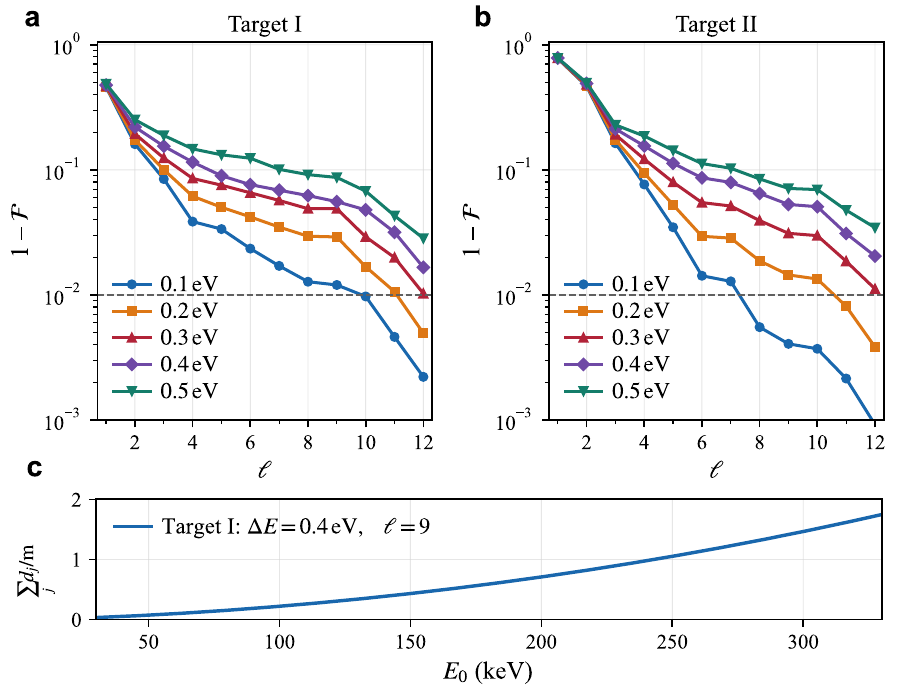}
\caption{
    PINEMAD state preparation with finite energy spread and required drift lengths. (a),(b) Optimized infidelity $1-\mathcal F_{\sigma_E}$ versus the number of PINEMAD stages $\ell$ for the finite-width counterparts of Targets I and II in Fig.~\hyperref[fig:2]{2}, respectively. Different colored lines with markers represent different incident energy widths $\Delta E$ (FWHM); the dashed line marks $\mathcal F_{\sigma_E}=99\%$. The calculations use an unchirped Gaussian incident wavepacket and an optical wavelength of $1030\,\mathrm{nm}$, with PINEMAD parameters optimized independently for each $\Delta E$ and $\ell$. (c) Total free-drift length $\sum_j L_j$ versus the central electron kinetic energy $E_0$ for Target I with $\Delta E=0.4\,\mathrm{eV}$ and $\ell=9$. This sequence has the largest total length among those in (a),(b).
    }
    \label{fig:5}
\end{figure}

The proof is provided in Sec.~V of the {Supplementary Information}. This corollary covers state preparation, unitary gates and isometric encodings, guaranteeing arbitrary accuracy for sufficiently small energy spread. To demonstrate state preparation at experimentally relevant energy widths $\Delta E=0.1$--$0.5\,\mathrm{eV}$, we reoptimize the sequences for Targets I and II, restricting the allowed drift phases and allowing $|g_j|\leq3$. The resulting fidelities as functions of $\ell$ and $\Delta E$ are shown in Fig.~\hyperref[fig:5]{5(a)--(b)}. With $\ell=12$, both targets achieve fidelities above $99\%$ at $\Delta E=0.1$ and $0.2\,\mathrm{eV}$, and above $96.5\%$ at $\Delta E=0.5\,\mathrm{eV}$. In experiments, the incident-electron energy spread can therefore be measured first and then used to optimize the finite-width PINEMAD sequence.

\textit{Effect of electron kinetic energy.}---The central electron kinetic energy $E_0$ affects PINEMAD in three ways. First, the separation between optical photon energies of a few $\mathrm{eV}$ and electron kinetic energies of tens to hundreds of $\mathrm{keV}$ supports the approximations in Eq.~\hyperref[operators]{(1)}, provided that the populated energy range is small compared with $E_0$ and dispersion during each interaction is negligible (see Sec.~I of the {Supplementary Information}). Second, $E_0$ influences the light--electron coupling strength; experimentally accessible couplings of order unity~\cite{henkeIntegratedPhotonicsEnables2021} are consistent with our optimization ranges. Third, $E_0$ determines the drift length required for a prescribed phase $\theta_j$:
\begin{equation}
    L_j=\frac{\theta_j}{2\pi}L_T,\qquad L_T=\frac{4\pi m_e\gamma_0^3v_0^3}{\hbar\omega_0^2},
\end{equation}
where $L_T$ is the electron Talbot length, $m_e$ is the electron rest mass, $\gamma_0=1+E_0/(m_ec^2)$, and $v_0=c\sqrt{1-\gamma_0^{-2}}$. For reference, Fig.~\hyperref[fig:5]{5(c)} shows the total drift length for the sequence with the largest total drift phase among those in Fig.~\hyperref[fig:5]{5(a)-(b)}, corresponding to Target I with $\Delta E=0.4\,\mathrm{eV}$ and $\ell=9$. At an optical wavelength of $1030\,\mathrm{nm}$, its total drift length increases from $0.033$ to $1.748\,\mathrm{m}$ as $E_0$ increases from $30$ to $330\,\mathrm{keV}$.

\textit{Summary.}---We have established that alternating PINEM interactions and free drift enable universal coherent control of free-electron energy sidebands. A finite PINEMAD sequence can approximate any prescribed isometry on a finite-dimensional input subspace, encompassing state preparation, unitary gates, and isometric encoding. High-fidelity state preparation remains achievable at realistic finite electron energy spreads with suitably chosen PINEMAD parameters and at electron kinetic energies commonly used in electron microscopy. The resulting freedom in electron preparation and measurement could support the heralded preparation of photonic states~\cite{dahanCreationOpticalCat2023,giulioTunableQuantumLight2025,huangElectronPhotonQuantumState2023,zhuProgrammingCoherentQuantum2026} and motivate future extensions of electron-heralded atomic-state preparation schemes~\cite{zhaoQuantumEntanglementModulation2021,synanidisRydbergAtomManipulationStrong2025,shenHeraldedFreeElectronWriting2026}. Future directions include determining how the optimal approximation error depends on the number of PINEMAD stages, and quantifying sensitivity to experimental control errors.

\textit{Acknowledgments.}---M.X. acknowledges support from the Key Research and Development Program of the Ministry of Science and Technology (Grant No. 2024YFB2808200), the National Natural Science Foundation of China (Grant Nos. 12321161645, 12334015, and 12274332), and the National Key Research and Development Program of China (Grant No. 2022YFA1404900). Z.Z. acknowledges support from the Alexander von Humboldt Foundation through a Postdoctoral Fellowship.

\textit{Author contributions.}---T.S. developed the theoretical framework, performed the analytical derivations and numerical calculations, and wrote the initial manuscript. Z.Z. and M.X. supervised the research, provided feedback, and revised the manuscript.

\textit{Competing interests.}---The authors declare no competing interests.

\textit{Data availability.}---The optimized PINEMAD sequence parameters and fidelity data supporting Figs.~2--5 are provided in the accompanying Supplementary Data file \texttt{PINEMAD\_parameters.csv}. The data definitions and procedures for reproducing the fidelities and drift lengths are given in Sec.~VI of the Supplementary Information.

\textit{Code availability.}---The numerical optimization and verification code used in this study is available from the corresponding authors upon reasonable request.

\bibliographystyle{apsrev4-2}
\bibliography{main}

@article{arquelopezAtomicFloquetPhysics2022,
  title = {Atomic {Floquet} physics revealed by free electrons},
  author = {{Arqu{\'{e}} L{\'{o}}pez}, Eduardo and {Di Giulio}, Valerio and {Garc{\'{i}}a de Abajo}, F. Javier},
  journal = {Phys. Rev. Research},
  volume = {4},
  pages = {013241},
  year = {2022},
  doi = {10.1103/physrevresearch.4.013241}
}

@article{barwickPhotoninducedNearfieldElectron2009,
  title = {Photon-induced near-field electron microscopy},
  author = {Barwick, Brett and Flannigan, David J. and Zewail, Ahmed H.},
  journal = {Nature},
  volume = {462},
  pages = {902--906},
  year = {2009},
  doi = {10.1038/nature08662}
}

@article{baumQuantumDynamicsAttosecond2017,
  title = {Quantum dynamics of attosecond electron pulse compression},
  author = {Baum, Peter},
  journal = {J. Appl. Phys.},
  volume = {122},
  pages = {223105},
  year = {2017},
  doi = {10.1063/1.5006864}
}

@article{braimanSyntheticHilbertSpace2023,
  title = {The Synthetic Hilbert Space of Laser-Driven Free-Electrons},
  author = {Braiman, Guy and Reinhardt, Ori and Mechel, Chen and Levi, Omer and Kaminer, Ido},
  journal = {Quantum},
  volume = {7},
  pages = {888},
  year = {2023},
  doi = {10.22331/q-2023-01-03-888}
}

@article{dahanCreationOpticalCat2023,
  title = {Creation of Optical Cat and {GKP} States Using Shaped Free Electrons},
  author = {Dahan, Raphael and Baranes, Gefen and Gorlach, Alexey and Ruimy, Ron and Rivera, Nicholas and Kaminer, Ido},
  journal = {Phys. Rev. X},
  volume = {13},
  pages = {031001},
  year = {2023},
  doi = {10.1103/physrevx.13.031001}
}

@article{feistQuantumCoherentOptical2015,
  title = {Quantum coherent optical phase modulation in an ultrafast transmission electron microscope},
  author = {Feist, Armin and Echternkamp, Katharina E. and Schauss, Jakob and Yalunin, Sergey V. and Sch{\"{a}}fer, Sascha and Ropers, Claus},
  journal = {Nature},
  volume = {521},
  pages = {200--203},
  year = {2015},
  doi = {10.1038/nature14463}
}

@article{garciadeabajoProbingPhotonicLocal2008,
  title = {Probing the Photonic Local Density of States with Electron Energy Loss Spectroscopy},
  author = {{Garc{\'{i}}a de Abajo}, F. J. and Kociak, M.},
  journal = {Phys. Rev. Lett.},
  volume = {100},
  pages = {106804},
  year = {2008},
  doi = {10.1103/physrevlett.100.106804}
}

@article{garciadeabajoRoadmapQuantumNanophotonics2025,
  title = {Roadmap for Quantum Nanophotonics with Free Electrons},
  author = {
    {Garc{\'i}a de Abajo}, F. Javier
    and Polman, Albert
    and Velasco, Cruz I.
    and Kociak, Mathieu
    and Tizei, Luiz H. G.
    and St{\'e}phan, Odile
    and Meuret, Sophie
    and Sannomiya, Takumi
    and Akiba, Keiichirou
    and Auad, Yves
    and others
    },
  journal = {ACS Photonics},
  volume = {12},
  pages = {4760--4817},
  year = {2025},
  doi = {10.1021/acsphotonics.5c00585}
}

@article{giulioTunableQuantumLight2025,
  title = {Tunable quantum light by modulated free electrons},
  author = {{Di Giulio}, Valerio and Haindl, Rudolf and Ropers, Claus},
  journal = {Nanophotonics},
  volume = {14},
  pages = {1865--1878},
  year = {2025},
  doi = {10.1515/nanoph-2025-0040}
}

@article{huangElectronPhotonQuantumState2023,
  title = {Electron-Photon Quantum State Heralding Using Photonic Integrated Circuits},
  author = {Huang, Guanhao and Engelsen, Nils J. and Kfir, Ofer and Ropers, Claus and Kippenberg, Tobias J.},
  journal = {PRX Quantum},
  volume = {4},
  pages = {020351},
  year = {2023},
  doi = {10.1103/prxquantum.4.020351}
}

@article{kfirOpticalCoherenceTransfer2021,
  title = {Optical coherence transfer mediated by free electrons},
  author = {Kfir, Ofer and {Di Giulio}, Valerio and {de Abajo}, F. Javier Garc{\'{i}}a and Ropers, Claus},
  journal = {Sci. Adv.},
  volume = {7},
  pages = {eabf6380},
  year = {2021},
  doi = {10.1126/sciadv.abf6380}
}

@article{konecnaEntanglingFreeElectrons2022,
  title = {Entangling free electrons and optical excitations},
  author = {Kone{\v{c}}n{\'{a}}, Andrea and Iyikanat, Fadil and {Garc{\'{i}}a de Abajo}, F. Javier},
  journal = {Sci. Adv.},
  volume = {8},
  pages = {eabo7853},
  year = {2022},
  doi = {10.1126/sciadv.abo7853}
}

@article{limQuantumInterferenceFundamentally2023,
  title = {Quantum Interference between Fundamentally Different Processes Is Enabled by Shaped Input Wavefunctions},
  author = {Lim, Jeremy and Kumar, Suraj and Ang, Yee Sin and Ang, Lay Kee and Wong, Liang Jie},
  journal = {Adv. Sci.},
  volume = {10},
  pages = {2205750},
  year = {2023},
  doi = {10.1002/advs.202205750}
}

@article{morimotoDiffractionMicroscopyAttosecond2018,
  title = {Diffraction and microscopy with attosecond electron pulse trains},
  author = {Morimoto, Yuya and Baum, Peter},
  journal = {Nat. Phys.},
  volume = {14},
  pages = {252--256},
  year = {2018},
  doi = {10.1038/s41567-017-0007-6}
}

@article{parkPhotoninducedNearfieldElectron2010,
  title = {Photon-induced near-field electron microscopy ({PINEM}): theoretical and experimental},
  author = {Park, Sang Tae and Lin, Milo and Zewail, Ahmed H},
  journal = {New J. Phys.},
  volume = {12},
  pages = {123028},
  year = {2010},
  doi = {10.1088/1367-2630/12/12/123028}
}

@article{polmanElectronbeamSpectroscopyNanophotonics2019a,
  title = {Electron-beam spectroscopy for nanophotonics},
  author = {Polman, Albert and Kociak, Mathieu and {Garc{\'{i}}a de Abajo}, F. Javier},
  journal = {Nat. Mater.},
  volume = {18},
  pages = {1158--1171},
  year = {2019},
  doi = {10.1038/s41563-019-0409-1}
}

@article{priebeAttosecondElectronPulse2017,
  title = {Attosecond electron pulse trains and quantum state reconstruction in ultrafast transmission electron microscopy},
  author = {Priebe, Katharina E. and Rathje, Christopher and Yalunin, Sergey V. and Hohage, Thorsten and Feist, Armin and Sch{\"{a}}fer, Sascha and Ropers, Claus},
  journal = {Nat. Photonics},
  volume = {11},
  pages = {793--797},
  year = {2017},
  doi = {10.1038/s41566-017-0045-8}
}

@article{reinhardtFreeElectronQubits2021,
  title = {Free-Electron Qubits},
  author = {Reinhardt, Ori and Mechel, Chen and Lynch, Morgan and Kaminer, Ido},
  journal = {Ann. Phys. (Berlin)},
  volume = {533},
  pages = {2000254},
  year = {2021},
  doi = {10.1002/andp.202000254}
}

@article{ruimyAtomicResolutionQuantumMeasurements2021,
  title = {Toward Atomic-Resolution Quantum Measurements with Coherently Shaped Free Electrons},
  author = {Ruimy, Ron and Gorlach, Alexey and Mechel, Chen and Rivera, Nicholas and Kaminer, Ido},
  journal = {Phys. Rev. Lett.},
  volume = {126},
  pages = {233403},
  year = {2021},
  doi = {10.1103/physrevlett.126.233403}
}

@article{ruimyFreeelectronQuantumOptics2025,
  title = {Free-electron quantum optics},
  author = {Ruimy, Ron and Karnieli, Aviv and Kaminer, Ido},
  journal = {Nat. Phys.},
  volume = {21},
  pages = {193--200},
  year = {2025},
  doi = {10.1038/s41567-024-02743-2}
}

@article{tsarevFreeelectronQubitsMaximumcontrast2021,
  title = {Free-electron qubits and maximum-contrast attosecond pulses via temporal {Talbot} revivals},
  author = {Tsarev, M. V. and Ryabov, A. and Baum, P.},
  journal = {Phys. Rev. Research},
  volume = {3},
  pages = {043033},
  year = {2021},
  doi = {10.1103/physrevresearch.3.043033}
}

@article{wongControlQuantumElectrodynamical2021,
  title = {Control of quantum electrodynamical processes by shaping electron wavepackets},
  author = {Wong, Liang Jie and Rivera, Nicholas and Murdia, Chitraang and Christensen, Thomas and Joannopoulos, John D. and Solja{\v{c}}i{\'{c}}, Marin and Kaminer, Ido},
  journal = {Nat. Commun.},
  volume = {12},
  pages = {1700},
  year = {2021},
  doi = {10.1038/s41467-021-21367-1}
}

@article{yaluninTailoredHighcontrastAttosecond2021,
  title = {Tailored high-contrast attosecond electron pulses for coherent excitation and scattering},
  author = {Yalunin, Sergey V. and Feist, Armin and Ropers, Claus},
  journal = {Phys. Rev. Research},
  volume = {3},
  pages = {L032036},
  year = {2021},
  doi = {10.1103/physrevresearch.3.l032036}
}

@article{zhangQuantumStateInterrogation2022,
  title = {Quantum state interrogation using a preshaped free electron wavefunction},
  author = {Zhang, Bin and Ran, Du and Ianconescu, Reuven and Friedman, Aharon and Scheuer, Jacob and Yariv, Amnon and Gover, Avraham},
  journal = {Phys. Rev. Research},
  volume = {4},
  pages = {033071},
  year = {2022},
  doi = {10.1103/physrevresearch.4.033071}
}

@article{zhaoQuantumEntanglementModulation2021,
  title = {Quantum Entanglement and Modulation Enhancement of Free-Electron--Bound-Electron Interaction},
  author = {Zhao, Zhexin and Sun, Xiao-Qi and Fan, Shanhui},
  journal = {Phys. Rev. Lett.},
  volume = {126},
  pages = {233402},
  year = {2021},
  doi = {10.1103/physrevlett.126.233402}
}

@misc{zhuProgrammingCoherentQuantum2026,
  title = {Programming Coherent and Quantum Light with a Free-Electron Wavepacket},
  author = {Zhu, Songyu and Zeng, Yushan and Pan, Chenhao and Pan, Yiming and Tian, Ye and Li, Ruxin},
  year = {2026},
  eprint = {2604.21246},
  archivePrefix = {arXiv},
  primaryClass = {physics.optics},
  doi = {10.48550/arXiv.2604.21246}
}

@article{karnieliQuantumSensingStrongly2023,
  title = {Quantum sensing of strongly coupled light-matter systems using free electrons},
  author = {Karnieli, Aviv and Tsesses, Shai and Yu, Renwen and Rivera, Nicholas and Zhao, Zhexin and Arie, Ady and Fan, Shanhui and Kaminer, Ido},
  journal = {Sci. Adv.},
  volume = {9},
  pages = {eadd2349},
  year = {2023},
  doi = {10.1126/sciadv.add2349}
}

@article{baranesFreeElectronsInduce2022,
  title = {Free electrons can induce entanglement between photons},
  author = {Baranes, Gefen and Ruimy, Ron and Gorlach, Alexey and Kaminer, Ido},
  journal = {npj Quantum Inf.},
  volume = {8},
  pages = {32},
  year = {2022},
  doi = {10.1038/s41534-022-00540-4}
}

@article{baranesFreeelectronInteractionsPhotonic2023,
  title = {Free-electron interactions with photonic {GKP} states: Universal control and quantum error correction},
  author = {Baranes, Gefen and Even-Haim, Shiran and Ruimy, Ron and Gorlach, Alexey and Dahan, Raphael and Diringer, Asaf A. and Hacohen-Gourgy, Shay and Kaminer, Ido},
  journal = {Phys. Rev. Research},
  volume = {5},
  pages = {043271},
  year = {2023},
  doi = {10.1103/physrevresearch.5.043271}
}

@article{wongFreeelectronCrystalsEnhanced2024,
  title = {Free-electron crystals for enhanced {X-ray} radiation},
  author = {Wong, Lee Wei Wesley and Shi, Xihang and Karnieli, Aviv and Lim, Jeremy and Kumar, Suraj and Carbajo, Sergio and Kaminer, Ido and Wong, Liang Jie},
  journal = {Light Sci. Appl.},
  volume = {13},
  pages = {29},
  year = {2024},
  doi = {10.1038/s41377-023-01363-4}
}

@article{henkeIntegratedPhotonicsEnables2021,
  title = {Integrated photonics enables continuous-beam electron phase modulation},
  author = {Henke, Jan-Wilke and Raja, Arslan Sajid and Feist, Armin and Huang, Guanhao and Arend, Germaine and Yang, Yujia and Kappert, F. Jasmin and Wang, Rui Ning and M{\"{o}}ller, Marcel and Pan, Jiahe and Liu, Junqiu and Kfir, Ofer and Ropers, Claus and Kippenberg, Tobias J.},
  journal = {Nature},
  volume = {600},
  pages = {653--658},
  year = {2021},
  doi = {10.1038/s41586-021-04197-5}
}

@article{auadMeVElectronSpectromicroscopy2023,
  title = {$\mu$eV electron spectromicroscopy using free-space light},
  author = {Auad, Yves and Dias, Eduardo J. C. and Tenc{\'{e}}, Marcel and Blazit, Jean-Denis and Li, Xiaoyan and Zagonel, Luiz Fernando and St{\'{e}}phan, Odile and Tizei, Luiz H. G. and {Garc{\'{i}}a de Abajo}, F. Javier and Kociak, Mathieu},
  journal = {Nat. Commun.},
  volume = {14},
  pages = {4442},
  year = {2023},
  doi = {10.1038/s41467-023-39979-0}
}

@article{dahanResonantPhasematchingLight2020,
  title = {Resonant phase-matching between a light wave and a free-electron wavefunction},
  author = {Dahan, Raphael and Nehemia, Saar and Shentcis, Michael and Reinhardt, Ori and Adiv, Yuval and Shi, Xihang and Be'er, Orr and Lynch, Morgan H. and Kurman, Yaniv and Wang, Kangpeng and Kaminer, Ido},
  journal = {Nat. Phys.},
  volume = {16},
  pages = {1123--1131},
  year = {2020},
  doi = {10.1038/s41567-020-01042-w}
}

@article{boscainMultiinputSchrodingerEquation2014,
  title = {Multi-input {Schr{\"{o}}dinger} equation: Controllability, tracking, and application to the quantum angular momentum},
  author = {Boscain, Ugo and Caponigro, Marco and Sigalotti, Mario},
  journal = {J. Differ. Equations},
  volume = {256},
  pages = {3524--3551},
  year = {2014},
  doi = {10.1016/j.jde.2014.02.004}
}

@misc{alvarez-ahedoPhaseProgrammableFreeElectron2026,
  title = {Phase-Programmable Free Electron Quantum States in Synthetic Momentum Space},
  author = {Alvarez-Ahedo, Alatz and Lazo, Miriam and Xu, Tian-Niu and Pan, Yiming and Sanz, Mikel and Ding, Yongcheng},
  year = {2026},
  eprint = {2607.07445},
  archivePrefix = {arXiv},
  primaryClass = {quant-ph},
  doi = {10.48550/arXiv.2607.07445}
}

@article{shilohQuantumCoherentLightElectronInteraction2022,
  title = {Quantum-Coherent Light-Electron Interaction in a Scanning Electron Microscope},
  author = {Shiloh, R. and Chlouba, T. and Hommelhoff, P.},
  journal = {Phys. Rev. Lett.},
  volume = {128},
  pages = {235301},
  year = {2022},
  doi = {10.1103/PhysRevLett.128.235301}
}

@article{wuCoherentControlThreeLevel2026,
  title = {Coherent Control of Three-Level System Using Shaped Free Electrons},
  author = {Wu, Dixuan and Li, Jing and Jiang, Yuhan and Liu, Yunquan},
  journal = {Phys. Rev. Lett.},
  volume = {137},
  pages = {063601},
  year = {2026},
  doi = {10.1103/l2y4-tpzb}
}

@misc{shenHeraldedFreeElectronWriting2026,
  title = {Heralded Free-Electron Writing of the Most Subradiant State in an Atomic Array},
  author = {Shen, Tong and Zhao, Zhexin and Xiao, Meng},
  year = {2026},
  eprint = {2608.03056},
  archivePrefix = {arXiv},
  primaryClass = {quant-ph},
  doi = {10.48550/arXiv.2608.03056}
}

@article{echternkampRamseytypePhaseControl2016,
  title = {Ramsey-type phase control of free-electron beams},
  author = {Echternkamp, Katharina E. and Feist, Armin and Sch{\"{a}}fer, Sascha and Ropers, Claus},
  journal = {Nat. Phys.},
  volume = {12},
  pages = {1000--1004},
  year = {2016},
  doi = {10.1038/nphys3844}
}

@article{reinhardtTheoryShapingElectron2020a,
  title = {Theory of Shaping Electron Wavepackets with Light},
  author = {Reinhardt, Ori and Kaminer, Ido},
  journal = {ACS Photonics},
  volume = {7},
  pages = {2859--2870},
  year = {2020},
  doi = {10.1021/acsphotonics.0c01133}
}

@article{vanacoreAttosecondCoherentControl2018,
  title = {Attosecond coherent control of free-electron wave functions using semi-infinite light fields},
  author = {Vanacore, G. M. and Madan, I. and Berruto, G. and Wang, K. and Pomarico, E. and Lamb, R. J. and McGrouther, D. and Kaminer, I. and Barwick, B. and {Garc{\'{i}}a de Abajo}, F. Javier and Carbone, F.},
  journal = {Nat. Commun.},
  volume = {9},
  pages = {2694},
  year = {2018},
  doi = {10.1038/s41467-018-05021-x}
}

@article{abad-arredondoQuantumStatePreparation2025,
  title = {Quantum state preparation and readout with modulated electrons},
  author = {Abad-Arredondo, J. and Fern{\'{a}}ndez-Dom{\'{i}}nguez, A. I.},
  journal = {Phys. Rev. B},
  volume = {111},
  pages = {165422},
  year = {2025},
  doi = {10.1103/physrevb.111.165422}
}

@article{feistCavitymediatedElectronphotonPairs2022,
  title = {Cavity-mediated electron-photon pairs},
  author = {Feist, Armin and Huang, Guanhao and Arend, Germaine and Yang, Yujia and Henke, Jan-Wilke and Raja, Arslan Sajid and Kappert, F. Jasmin and Wang, Rui Ning and Louren{\c{c}}o-Martins, Hugo and Qiu, Zheru and Liu, Junqiu and Kfir, Ofer and Kippenberg, Tobias J. and Ropers, Claus},
  journal = {Science},
  volume = {377},
  pages = {777--780},
  year = {2022},
  doi = {10.1126/science.abo5037}
}

@article{arendElectronsHeraldNonclassical2025,
  title = {Electrons herald non-classical light},
  author = {Arend, Germaine and Huang, Guanhao and Feist, Armin and Yang, Yujia and Henke, Jan-Wilke and Qiu, Zheru and Jeng, Hao and Raja, Arslan Sajid and Haindl, Rudolf and Wang, Rui Ning and Kippenberg, Tobias J. and Ropers, Claus},
  journal = {Nat. Phys.},
  volume = {21},
  pages = {1855--1862},
  year = {2025},
  doi = {10.1038/s41567-025-03033-1}
}

@article{garciadeabajoMultiphotonAbsorptionEmission2010a,
  title = {Multiphoton Absorption and Emission by Interaction of Swift Electrons with Evanescent Light Fields},
  author = {{Garc{\'{i}}a de Abajo}, F. Javier and Asenjo-Garcia, Ana and Kociak, Mathieu},
  journal = {Nano Lett.},
  volume = {10},
  pages = {1859--1863},
  year = {2010},
  doi = {10.1021/nl100613s}
}

@article{giulioFreeelectronShapingUsing2020,
  title = {Free-electron shaping using quantum light},
  author = {{Di Giulio}, Valerio and {Garc{\'{i}}a de Abajo}, F. Javier},
  journal = {Optica},
  volume = {7},
  pages = {1820--1830},
  year = {2020},
  doi = {10.1364/optica.404598}
}

@article{kfirEntanglementsElectronsCavity2019b,
  title = {Entanglements of Electrons and Cavity Photons in the Strong-Coupling Regime},
  author = {Kfir, Ofer},
  journal = {Phys. Rev. Lett.},
  volume = {123},
  pages = {103602},
  year = {2019},
  doi = {10.1103/physrevlett.123.103602}
}

@article{dahanImprintingQuantumStatistics2021,
  title = {Imprinting the quantum statistics of photons on free electrons},
  author = {Dahan, Raphael and Gorlach, Alexey and Haeusler, Urs and Karnieli, Aviv and Eyal, Ori and Yousefi, Peyman and Segev, Mordechai and Arie, Ady and Eisenstein, Gadi and Hommelhoff, Peter and Kaminer, Ido},
  journal = {Science},
  volume = {373},
  pages = {eabj7128},
  year = {2021},
  doi = {10.1126/science.abj7128}
}

@article{pescollerCouplingFreeElectrons2026,
  title = {Coupling free electrons to a trapped-ion quantum computer},
  author = {Pescoller, Elias and Beltr{\'{a}}n-Romero, Santiago and Egginger, Sebastian and Jungwirth, Nicolas and Zanetti, Martino and Hornof, Dominik and Seifner, Michael S. and B{\v{r}}ezinov{\'{a}}, Iva and Haslinger, Philipp and Juffmann, Thomas and Kofler, Johannes and R{\"{a}}tzel, Dennis and Schindler, Philipp},
  journal = {Phys. Rev. Lett.},
  year = {2026},
  note = {accepted},
  doi = {10.1103/w6t7-9txs}
}

@article{garciadeabajoOpticalModulationElectron2021,
  title = {Optical Modulation of Electron Beams in Free Space},
  author = {{Garc{\'{i}}a de Abajo}, F. Javier and Kone{\v{c}}n{\'{a}}, Andrea},
  journal = {Phys. Rev. Lett.},
  volume = {126},
  pages = {123901},
  year = {2021},
  doi = {10.1103/PhysRevLett.126.123901}
}

@article{goverFreeElectronBoundElectronResonant2020,
  title = {Free-Electron--Bound-Electron Resonant Interaction},
  author = {Gover, Avraham and Yariv, Amnon},
  journal = {Phys. Rev. Lett.},
  volume = {124},
  pages = {064801},
  year = {2020},
  doi = {10.1103/physrevlett.124.064801}
}

@article{parkRelativisticEffectsPhotonInduced2012,
  title = {Relativistic Effects in Photon-Induced Near Field Electron Microscopy},
  author = {Park, Sang Tae and Zewail, Ahmed H.},
  journal = {J. Phys. Chem. A},
  volume = {116},
  pages = {11128--11133},
  year = {2012},
  doi = {10.1021/jp304534n}
}

@article{trotterProductSemigroups1959,
  title = {On the Product of Semi-Groups of Operators},
  author = {Trotter, H. F.},
  journal = {Proc. Am. Math. Soc.},
  volume = {10},
  pages = {545--551},
  year = {1959},
  doi = {10.1090/S0002-9939-1959-0108732-6}
}

@article{synanidisRydbergAtomManipulationStrong2025,
  title = {{Rydberg-Atom Manipulation through Strong Interaction with Free Electrons}},
  author = {Synanidis, Adamantios P. and Gon{\c c}alves, P. A. D. and Garc{\'i}a de Abajo, F. Javier},
  journal = {ACS Nano},
  volume = {19},
  number = {12},
  pages = {11891--11899},
  year = {2025},
  doi = {10.1021/acsnano.4c14658}
}

@article{giulioModulationCathodoluminescence2021,
  title = {Modulation of Cathodoluminescence Emission by Interference with External Light},
  author = {Di Giulio, Valerio and Kfir, Ofer and Ropers, Claus and Garc{\'i}a de Abajo, F. Javier},
  journal = {ACS Nano},
  volume = {15},
  pages = {7290--7304},
  year = {2021},
  doi = {10.1021/acsnano.1c00549}
}

@article{morimotoCoherentScatteringOptically2021,
  title = {Coherent scattering of an optically modulated electron beam by atoms},
  author = {Morimoto, Yuya and Hommelhoff, Peter and Madsen, Lars Bojer},
  journal = {Phys. Rev. A},
  volume = {103},
  pages = {043110},
  year = {2021},
  doi = {10.1103/PhysRevA.103.043110}
}

@article{garciadeabajoSpatiotemporalElectronBeam2023,
  title = {Spatiotemporal Electron Beam Focusing through Parallel Interactions with Shaped Optical Fields},
  author = {Garc{\'i}a de Abajo, F. Javier and Ropers, Claus},
  journal = {Phys. Rev. Lett.},
  volume = {130},
  pages = {246901},
  year = {2023},
  doi = {10.1103/PhysRevLett.130.246901}
}

@article{zhaoElectronPulseCompression2021,
  title = {Electron Pulse Compression with Optical Beat Note},
  author = {Zhao, Zhexin and Leedle, Kenneth J. and Black, Dylan S. and Solgaard, Olav and Byer, Robert L. and Fan, Shanhui},
  journal = {Phys. Rev. Lett.},
  volume = {127},
  pages = {164802},
  year = {2021},
  doi = {10.1103/PhysRevLett.127.164802}
}

@article{krivanekVibrationalSpectroscopyElectron2014,
  title = {Vibrational spectroscopy in the electron microscope},
  author = {Krivanek, Ondrej L. and Lovejoy, Tracy C. and Dellby, Niklas and Aoki, Toshihiro and Carpenter, R. W. and Rez, Peter and Soignard, Emmanuel and Zhu, Jiangtao and Batson, Philip E. and Lagos, Maureen J. and Egerton, Ray F. and Crozier, Peter A.},
  journal = {Nature},
  volume = {514},
  pages = {209--212},
  year = {2014},
  doi = {10.1038/nature13870}
}

\end{document}


\title{{Supplementary Information} for \textquotedblleft Universal Free-Electron-Sideband Control\textquotedblright}
\author{Tong Shen}
\affiliation{School of Physics, Wuhan University, Wuhan 430072, China}
\author{Zhexin Zhao}
\email{zhexin.zhao@fau.de}
\affiliation{Department of Physics, Friedrich-Alexander University (FAU) Erlangen--N\"urnberg, Staudtstra\ss e 1, 91058 Erlangen, Germany}
\author{Meng Xiao}
\email{phmxiao@whu.edu.cn}
\affiliation{School of Physics, Wuhan University, Wuhan 430072, China}
\affiliation{Wuhan Institute of Quantum Technology, Wuhan 430206, China}

\maketitle
\onecolumngrid
\vspace{5mm}

\setcounter{section}{0}
\setcounter{secnumdepth}{1}
\renewcommand{\thesection}{\Roman{section}}

\setcounter{equation}{0}
\renewcommand{\theequation}{S\arabic{equation}}

\setcounter{figure}{0}
\renewcommand{\thefigure}{S\arabic{figure}}

\setcounter{table}{0}
\renewcommand{\thetable}{S\arabic{table}}

\begingroup
\makeatletter
\let\l@subsection\@gobbletwo
\let\l@subsubsection\@gobbletwo
\let\l@paragraph\@gobbletwo
\let\l@subparagraph\@gobbletwo
\makeatother
\tableofcontents
\endgroup

\clearpage

\section{PINEM and free-drift operators}\label{sec:operators}

{We review the derivation of the PINEM and free-drift operators and the approximations underlying their use, establishing the notation used throughout this work.} For an electron wavepacket propagating along the $z$ direction, we adopt a slowly varying envelope description ~\cite{parkRelativisticEffectsPhotonInduced2012, garciadeabajoOpticalModulationElectron2021} and write
\begin{equation}
    \psi(\mathbf r,t) =\exp[\ii(p_0z-E_0t)/\hbar]\phi(\mathbf r,t),
\end{equation}
where $\mathbf r=(x,y,z)$, $E_0$ is the reference electron kinetic energy, and $p_0=m_e\gamma_0v_0$ is the reference longitudinal momentum. Here $m_e$ is the electron rest mass, $v_0$ is the reference velocity, and $\gamma_0=(1-v_0^2/c^2)^{-1/2}$ is the Lorentz factor. The constant rest energy is omitted because it contributes only an overall phase.

In a gauge with zero scalar potential, the envelope obeys the modified Schr\"odinger equation~\cite{garciadeabajoRoadmapQuantumNanophotonics2025}
\begin{equation}
    \label{eq:master-envelope}
    \ii\hbar(\partial_t+v_0\partial_z)\phi =-\frac{\hbar^2}{2m_e\gamma_0} \left(\partial_{xx}+\partial_{yy} +\gamma_0^{-2}\partial_{zz}\right)\phi +\mathcal H^{\mathrm{int}}\phi,
\end{equation}
with
\begin{equation}
    \mathcal H^{\mathrm{int}} =\frac{ev_0}{c}A_z +\frac{e^2}{2m_ec^2\gamma_0} \left(A_x^2+A_y^2+\frac{A_z^2}{\gamma_0^2}\right).
\end{equation}
The electron charge is $-e$, with $e>0$, and $A_x,A_y,A_z$ are the components of the external vector potential. We use $\partial_{xx}=\partial^2/\partial x^2$, with analogous notation for $y$ and $z$. Here $x$ and $y$ denote transverse spatial coordinates.

We focus on longitudinal dynamics and assume that the optical coupling is approximately uniform across the transverse electron profile. We further assume that the characteristic quadratic interaction energy is negligible compared with the linear coupling,
\begin{equation}
    \left| \frac{e^2}{2m_ec^2\gamma_0} \left(A_x^2+A_y^2+\frac{A_z^2}{\gamma_0^2}\right) \right|_{\mathrm{char}} \ll \left|\frac{ev_0}{c}A_z\right|_{\mathrm{char}}.
\end{equation}
Retaining only the linear interaction ~\cite{garciadeabajoMultiphotonAbsorptionEmission2010a, parkPhotoninducedNearfieldElectron2010, garciadeabajoRoadmapQuantumNanophotonics2025}, Eq.~(\ref{eq:master-envelope}) reduces to
\begin{equation}
    \label{eq:longitudinal_envelope}
    \ii\hbar(\partial_t+v_0\partial_z)\phi =-\frac{\hbar^2}{2m_e\gamma_0^3}\partial_{zz}\phi +\frac{ev_0}{c}A_z(z,t)\phi.
\end{equation}

\subsection{PINEM interaction}
During the PINEM interaction, we adopt the nonrecoil approximation, in which the electron dispersion is linearized around the reference momentum $p_0$:
\begin{equation}
    \label{eq:linear_dispersion}
    E(p_z)\simeq E_0+v_0(p_z-p_0).
\end{equation}
This approximation neglects the quadratic dispersion term $(p_z-p_0)^2/(2m_e\gamma_0^3)$ and hence the corresponding second spatial derivative in Eq.~(\ref{eq:longitudinal_envelope}). It is valid when the populated energy offsets remain small compared with $E_0$ and the dispersion phase accumulated over the interaction length is negligible, as quantified in Sec.~\hyperref[sec:pinem_approximations]{I.C}. The envelope equation therefore reduces to
\begin{equation}
    \ii\hbar(\partial_t+v\partial_z)\phi =\frac{ev}{c}A_z(z,t)\phi.
\end{equation}

In PINEM, the electron interacts with an optical near field. We consider a classical monochromatic field and write its longitudinal vector-potential component as
\begin{equation}
    \label{eq:monochromatic_interaction}
    A_z(z,t)=\mathcal A_z(z)\exp(-\ii\omega_0t) +\mathcal A_z^*(z)\exp(\ii\omega_0t),
\end{equation}
where $\omega_0$ is the optical angular frequency and $\mathcal A_z(z)$ is the complex vector-potential amplitude evaluated along the electron trajectory. To solve the envelope equation, we introduce the retarded-time coordinate $\tau=t-z/v_0$, which is constant along the reference electron trajectory. At fixed $\tau$, the equation becomes
\begin{equation}
    \ii\hbar v_0\,\partial_z\phi(z,\tau)=\frac{ev_0}{c}\Big[\mathcal A_z(z)\exp(-\ii\omega_0z/v_0)\exp(-\ii\omega_0\tau)+\mathcal A_z^*(z)\exp(\ii\omega_0z/v_0)\exp(\ii\omega_0\tau)\Big]\phi(z,\tau).
\end{equation}
Integration through the interaction region gives
\begin{equation}
    \label{eq:envelope_pinem}
    \begin{split}
        \phi_{\mathrm{out}}(\tau) &=\exp\left[g\exp(-\ii\omega_0\tau) -g^*\exp(\ii\omega_0\tau)\right]\phi_{\mathrm{in}}(\tau),\\
        g&=-\frac{\ii e}{\hbar c} \int\mathrm{d}z\,\mathcal A_z(z)\exp(-\ii\omega_0z/v_0).
    \end{split}
\end{equation}
Here $g$ is the generally complex, dimensionless PINEM coupling, and the integral extends over the interaction region. In the zero-scalar-potential gauge adopted above, the complex electric-field amplitude satisfies $\mathcal E_z(z)=\ii\omega_0\mathcal A_z(z)/c$. The coupling can therefore also be expressed as
\begin{equation}
    \label{eq:field_coupling}
    g=-\frac{e}{\hbar\omega_0} \int\mathrm{d}z\,\mathcal E_z(z)\exp(-\ii\omega_0z/v_0).
\end{equation}

To express the envelope modulation in the energy-sideband basis, we note that multiplication by $\exp(-\ii\omega_0\tau) =\exp(-\ii\omega_0t)\exp(\ii\omega_0z/v_0)$ increases the electron energy by $\hbar\omega_0$ and its longitudinal momentum by $\hbar\omega_0/v_0$, consistently with the linear-dispersion approximation in Eq.~(\ref{eq:linear_dispersion}). This multiplication is therefore represented by the energy-raising operator $b^\dagger$, while its complex conjugate corresponds to the energy-lowering operator $b$ ~\cite{feistQuantumCoherentOptical2015, kfirEntanglementsElectronsCavity2019b}. In the ideal sideband Hilbert space $\mathcal H=\ell^2(\mathbb Z)$, with $\ket{n}\equiv\ket{E_0+n\hbar\omega_0}$, these operators and the sideband-index operator $N$ act as
\begin{equation}
    b\ket{n}=\ket{n-1},\qquad b^\dagger\ket{n}=\ket{n+1},\qquad N\ket{n}=n\ket{n}.
\end{equation}
Equation~(\ref{eq:envelope_pinem}) thus yields the PINEM operator,
\begin{equation}
    \label{eq:A}
    \ket{\psi_{\mathrm{out}}} =A(g)\ket{\psi_{\mathrm{in}}},\qquad A(g)=\exp(gb^\dagger-g^*b).
\end{equation}
On this ladder, $bb^\dagger=b^\dagger b=\mathds1$, so $[b,b^\dagger]=0$. For $g\neq0$, the Jacobi--Anger expansion gives
\begin{equation}
    \label{eq:bessel}
    A(g)\ket{m} =\sum_{n\in\mathbb Z}J_{n-m}(2|g|) \exp[\ii(n-m)\arg g]\ket{n},
\end{equation}
where $J_q$ is the Bessel function of the first kind of integer order $q$. For $g=0$, $A(0)=\mathds1$.

Since $b$ and $b^\dagger$ commute, consecutive PINEM interactions without intervening drift satisfy
\begin{equation}
    A(g_2)A(g_1)=A(g_1+g_2).
\end{equation}
Thus, any sequence consisting solely of PINEM interactions is equivalent to a single interaction with the summed coupling.

\subsection{Free drift and its physical length}

During free drift, the optical interaction vanishes. Unlike during the short PINEM interaction, quadratic longitudinal dispersion can accumulate an appreciable phase over the drift distance and must therefore be retained. Setting $A_z=0$ in Eq.~(\ref{eq:longitudinal_envelope}) gives
\begin{equation}
    \label{eq:longitudinal_drift}
    \ii\hbar(\partial_t+v_0\partial_z)\phi =-\frac{\hbar^2}{2m_e\gamma_0^3}\partial_{zz}\phi.
\end{equation}
To determine the phase accumulated by an energy component, we consider an electron with kinetic energy $E_0+\epsilon$ and envelope $\phi_\epsilon(z,t)\propto \exp[\ii k(\epsilon)z-\ii\epsilon t/\hbar]$, where $\hbar k(\epsilon)$ is the longitudinal momentum offset from $p_0$. Substituting this envelope into Eq.~(\ref{eq:longitudinal_drift}) yields
\begin{equation}
    \epsilon=\hbar v_0k(\epsilon) +\frac{\hbar^2k(\epsilon)^2}{2m_e\gamma_0^3}.
\end{equation}
Expanding $k(\epsilon)$ to second order in $\epsilon$ gives
\begin{equation}
    k(\epsilon) =\frac{\epsilon}{\hbar v_0} -\frac{\epsilon^2}{2\hbar m_e\gamma_0^3v_0^3} +O(\epsilon^3).
\end{equation}
Propagation over a distance $L_j$ contributes the spatial phase $\exp[\ii k(\epsilon)L_j]$. To express the envelope in the retarded-time coordinate $\tau=t-z/v_0$, we write
\begin{equation}
    \exp[\ii k(\epsilon)z-\ii\epsilon t/\hbar] =\exp\left[\ii\left(k(\epsilon) -\frac{\epsilon}{\hbar v_0}\right)z\right] \exp(-\ii\epsilon\tau/\hbar).
\end{equation}
Thus, the linear propagation phase is absorbed into the retarded-time coordinate, leaving the quadratic phase
\begin{equation}
    \exp\left[-\frac{\ii L_j\epsilon^2} {2\hbar m_e\gamma_0^3v_0^3}\right].
\end{equation}
For an ideal sideband, $\epsilon=n\hbar\omega_0$, this phase becomes $\exp(-\ii\theta_jn^2)$. Free drift therefore acts on the sideband ladder as
\begin{equation}
    \label{eq:drift_length}
    U(\theta_j)=\exp(-\ii\theta_jN^2), \qquad \theta_j=\frac{L_j\hbar\omega_0^2} {2m_e\gamma_0^3v_0^3} =\frac{2\pi L_j}{L_T}, \qquad L_T=\frac{4\pi m_e\gamma_0^3v_0^3}{\hbar\omega_0^2}.
\end{equation}
Here $L_T$ is the electron Talbot length, at which all ideal sidebands recover their relative phases within the quadratic-dispersion approximation. The connection between longitudinal dispersion and electron pulse compression is discussed in Ref.~\cite{baumQuantumDynamicsAttosecond2017}.

On the ideal ladder, $U(\theta+2\pi)=U(\theta)$ because the sideband index $n$ is an integer. This periodicity does not generally hold for continuous energy offsets. Consequently, finite-width wavepacket calculations must retain the actual drift phases $\theta_j$, and physical drift lengths must be calculated from those phases rather than their values modulo $2\pi$.

\subsection{Validity of the PINEM approximations}
\label{sec:pinem_approximations}

The validity of these approximations depends on the energy range, interaction length, and field strength. The appreciably populated energy and momentum offsets must satisfy $|\epsilon|\ll E_0$ and $|\Delta p_z|\ll p_0$, so that the electron velocity remains close to $v_0$ and the dispersion expansion around the reference energy is valid.

Let $L_{\text{int}}$ be the interaction length and $\epsilon_{\max}$ the largest absolute energy offset with appreciable population during the interaction. Neglecting quadratic dispersion within the interaction region requires
\begin{equation}
    \label{eq:interaction_dispersion_energy}
    \frac{L_{\text{int}}\epsilon_{\max}^2} {2\hbar m_e\gamma_0^3v_0^3}\ll1.
\end{equation}
For the ideal sideband ladder, taking $\epsilon_{\max}=n_{\max}\hbar\omega_0$, where $n_{\max}$ is the largest absolute sideband index with appreciable population during the interaction, gives
\begin{equation}
    \label{eq:interaction_dispersion_condition}
    \frac{2\pi L_{\text{int}}}{L_T}\,n_{\max}^2\ll1.
\end{equation}
Under this condition, the PINEM interaction can be described using linear dispersion, while quadratic dispersion is retained over the subsequent free-drift interval.

The numerical examples use $|g_j|\leq2$ for ideal state preparation and unitary gates, $|g_j|\leq1$ for isometric encoding, and $|g_j|\leq3$ for finite-width state preparation. These coupling strengths lie within experimentally demonstrated ranges. Henke et al. report an interaction length of approximately $19\,\mu\mathrm m$, with the coupling strength controlled by the optical power~\cite{henkeIntegratedPhotonicsEnables2021}. Conventional localized near-field PINEM interactions typically extend over hundreds of nanometres, as discussed by Dahan et al. ~\cite{dahanResonantPhasematchingLight2020}. For comparison, Echternkamp et al. explicitly show that quadratic dispersion over a $6\,\mu\mathrm m$ propagation distance between two PINEM interactions is negligible for their experimental parameters ~\cite{echternkampRamseytypePhaseControl2016}. At the $1030\,\mathrm{nm}$ wavelength used in this work, Eq.~(\ref{eq:drift_length}) gives $L_T\simeq0.037$--$1.95\,\mathrm m$ for $E_0=30$--$330\,\mathrm{keV}$. These Talbot lengths are much larger than the localized interaction lengths, supporting the neglect of quadratic dispersion during PINEM when the populated sidebands satisfy Eq.~(\ref{eq:interaction_dispersion_condition}).

For a concrete example, take $L_{\text{int}}=1\,\mu\mathrm m$, $E_0=200\,\mathrm{keV}$, and $n_{\max}=20$ at the $1030\,\mathrm{nm}$ optical wavelength. This short interaction is compatible with the coupling strengths considered here: for a longitudinal field with uniform amplitude and constant spatial phase over the interaction region, Eq.~(\ref{eq:field_coupling}) gives a peak electric field of approximately $33\,\mathrm{MV/m}$ for $|g|=3$. This field amplitude is comparable to the $23$--$68\,\mathrm{MV/m}$ incident optical fields used in a coherent PINEM experiment ~\cite{feistQuantumCoherentOptical2015}. The comparison indicates similar field magnitudes, although the achievable longitudinal near field depends on the interaction geometry. Equation~(\ref{eq:drift_length}) gives $L_T\simeq0.792\,\mathrm m$, and the maximum quadratic-dispersion phase accumulated during the interaction is only $2\pi L_{\text{int}}n_{\max}^2/L_T \simeq3.2\times10^{-3}\,\mathrm{rad}$. Thus, quadratic dispersion can be neglected during this short PINEM interaction. At $E_0=30\,\mathrm{keV}$, the same wavelength, interaction length, and sideband range give a larger phase of approximately $6.8\times10^{-2}\,\mathrm{rad}$, indicating that the approximation requires closer scrutiny at lower electron energies.

\section{Proof of Theorem 1}\label{sec:theorem}

For any finite-dimensional input subspace $\mathcal E\subset\mathcal H$, linear inner-product-preserving map $T:\mathcal E\to\mathcal H$, and $\varepsilon>0$, we are going to prove that there is a finite PINEMAD sequence $W_\ell$ satisfying
\begin{equation}
    \label{eq:theorem_statement}
    \left\|(W_\ell-T)P_{\mathcal E}\right\|<\varepsilon.
\end{equation}
Here $P_{\mathcal E}$ is the orthogonal projector onto the input subspace $\mathcal E$, and $\|\cdot\|$ denotes the operator norm.

We are going to use Theorem~2.6 of Boscain, Caponigro, and Sigalotti~\cite{boscainMultiinputSchrodingerEquation2014}, abbreviated as BCS. Our proof first explains how this control result yields a finite PINEMAD sequence, then verifies that its hypotheses hold for our system. The three subsections have the following roles:
\begin{enumerate}
    \item \textit{Section II.A: state the BCS result and derive Theorem~1 conditionally.} We list the BCS hypotheses and the simultaneous-control conclusion used here. We show that Theorem~1 follows if these hypotheses hold for our system.
    \item \textit{Section II.B: verify the basic operator hypotheses.} We verify the basic operator assumptions required by BCS Theorem~2.6.
    \item \textit{Section II.C: verify the Lie--Galerkin Control Condition.} We verify the Lie--Galerkin Control Condition required by the same theorem.
\end{enumerate}

\subsection{BCS result and conditional derivation of Theorem 1}
\label{sec:bcs_result}
We first state the BCS result of Ref.~\cite{boscainMultiinputSchrodingerEquation2014}, writing its generators as $\mathsf A,\mathsf B_j$ and its allowed control sets as $\mathcal U,\mathcal U_j$ to distinguish them from the PINEMAD operators $A(g)$ and $U(\theta)$. Let $\mathcal H$ be an infinite-dimensional Hilbert space, and let $\mathsf A,\mathsf B_1,\ldots,\mathsf B_p$ be skew-adjoint (anti-Hermitian in physics terminology) operators. The controlled equation is
\begin{equation}
    \label{eq:bcs_abstract_dynamics}
    \frac{\mathrm d}{\mathrm dt}\psi(t)=\left[\mathsf A+\sum_{j=1}^{p}u_j(t)\mathsf B_j\right]\psi(t),\qquad u(t)\in\mathcal U.
\end{equation}
The operator $\mathsf A$ generates the evolution without applied controls and is called the drift generator. Each real-valued control $u_j(t)$ can be chosen independently within its allowed range $\mathcal U_j$. Thus $\mathcal U=\mathcal U_1\times\cdots\times\mathcal U_p$, where each $\mathcal U_j$ is either $[0,\delta]$ or $[-\delta,\delta]$ for a fixed $\delta>0$. Also, write $(\phi_k)_{k\in\mathbb N}$ for a complete orthonormal basis of drift eigenvectors, with $\mathsf A\phi_k=\mathrm{i}\lambda_k\phi_k$ and $\lambda_k\in\mathbb R$. The hypotheses of BCS Theorem~2.6 are:
\begin{enumerate}
    \item \textit{(A1), discrete drift spectrum.} The operator $\mathsf A$ has discrete spectrum with infinitely many distinct eigenvalues; degeneracies are allowed.
    \item \textit{(A2), control domains.} Each $\phi_k$ belongs to the domain of every $\mathsf B_j$.
    \item \textit{(A3), constant control generators.} For every allowed constant control vector $u\in\mathcal U$, the operator $\mathsf A+\sum_{j=1}^{p}u_j\mathsf B_j$ is essentially skew-adjoint (will be explained in Sect.~\hyperref[sec:bcs_operators]{II.B}) on the subspace
\begin{equation}
    \label{eq:bcs_finite_span}
    \mathcal L=\left\{\sum_{k\in\mathbb N}c_k\phi_k:c_k\neq0\text{ for only finitely many }k\right\}\subsetneq\mathcal H.
\end{equation}
    These three assumptions constitute condition $(A)$ in BCS Definition~2.1.
    \item \textit{Lie--Galerkin Control Condition.} This condition is defined in BCS Definition~2.5 and will be explained in detail and verified in Sec.~\hyperref[sec:bcs_lie]{II.C}.
\end{enumerate}
BCS Theorem~2.6 states that these hypotheses imply $L^1$-bounded approximate simultaneous controllability. We only use its weaker consequence, approximate simultaneous controllability, without the additional $L^1$ bound: for every finite family $\psi_1,\ldots,\psi_r\in\mathcal H$, target unitary $\widehat\Upsilon$, and $\varepsilon>0$, there are a finite time $T_0$ and piecewise-constant controls $u_j:[0,T_0]\to\mathcal U_j$ whose common propagator $\Upsilon_T^u$ satisfies
\begin{equation}
    \label{eq:bcs_simultaneous_control}
    \left\|\Upsilon_T^u\psi_k-\widehat\Upsilon\psi_k\right\|<\varepsilon,\qquad k=1,\ldots,r.
\end{equation}

We now map this abstract system to PINEMAD using the operators defined in Sec.~\hyperref[sec:operators]{I}. Take $p=2$ and, for convenience, $\delta=1$ (universality within this restricted control range is sufficient to establish universality with unrestricted $g\in\mathbb C$), and identify
\begin{equation}
    \label{eq:bcs_pinemad_correspondence}
    \begin{aligned}
        \mathsf A&=-\mathrm{i}N^2,\qquad\mathsf B_1=b^\dagger-b,\qquad\mathsf B_2=\mathrm{i}(b^\dagger+b),\\
        u_1(t)&=\operatorname{Re}g(t),\qquad u_2(t)=\operatorname{Im}g(t),\qquad\mathcal U=[-1,1]^2.
    \end{aligned}
\end{equation}
The abstract eigenvectors $\phi_k$ correspond to the sideband states $\ket n$ after choosing an enumeration of $n\in\mathbb Z$; the positive index $k$ labels the enumeration and is not the sideband index $n$. With this identification, Eq.~(\ref{eq:bcs_abstract_dynamics}) becomes
\begin{equation}
    \label{eq:auxiliary}
    \frac{\mathrm d}{\mathrm dt}\ket{\psi}=\left[-\mathrm{i}N^2+g(t)b^\dagger-g(t)^*b\right]\ket{\psi},\qquad\operatorname{Re}g(t),\operatorname{Im}g(t)\in[-1,1].
\end{equation}
Here $t$ is dimensionless and $g(t)$ is a piecewise-constant complex control. A constant $g$ acting for duration $\tau$ without drift gives $A(\tau g)$, so $g(t)$ specifies the coupling per unit of auxiliary time. This auxiliary equation allows drift and coupling to act simultaneously.

Temporarily assume that Eq.~(\ref{eq:auxiliary}) satisfies all four hypotheses, which we will verify in Secs.~\hyperref[sec:bcs_operators]{II.B} and~\hyperref[sec:bcs_lie]{II.C}. We first show that this is sufficient to obtain Eq.~(\ref{eq:theorem_statement}). Choose an orthonormal basis $\{\ket{e_a}\}_{a=1}^{d}$ of $\mathcal E$, where $d=\dim\mathcal E\geq1$. These basis states need not be individual sideband states $\ket n$; they may be sideband superpositions. Since $T$ preserves inner products, the output states $T\ket{e_a}$ are also orthonormal. Completing the input and output sets to orthonormal bases of $\mathcal H$ defines a unitary extension $\widehat T$ of $T$, which agrees with $T$ on $\mathcal E$.

In the BCS statement, choose $\widehat\Upsilon=\widehat T$ and apply Eq.~(\ref{eq:bcs_simultaneous_control}) to these $d$ basis states with tolerance $\varepsilon/(2\sqrt d)$. Denote the resulting common propagator by $V$ (a continuous-time evolution). If the control consists of $m$ constant segments with durations $\tau_j$ and values $g^{(j)}$, then
\begin{equation}
    \label{eq:continuous_propagator}
    \begin{split}
        V={}&\exp\left[\tau_m\bigl(-\mathrm{i}N^2+g^{(m)}b^\dagger-(g^{(m)})^*b\bigr)\right]\\
        &\cdots\exp\left[\tau_1\bigl(-\mathrm{i}N^2+g^{(1)}b^\dagger-(g^{(1)})^*b\bigr)\right].
    \end{split}
\end{equation}
The rightmost factor acts first. Here $g^{(j)}$ is the constant value of $g(t)$ on segment $j$, while $g_j$ in a PINEMAD sequence denotes the integrated coupling of one PINEM interaction. By construction, $V$ satisfies
\begin{equation}
    \left\|(V-T)\ket{e_a}\right\|<\frac{\varepsilon}{2\sqrt d},\qquad a=1,\ldots,d.
\end{equation}
Every normalized input in $\mathcal E$ is a superposition of these basis states. Cauchy--Schwarz bounds the error of any such superposition by the square root of the sum of the squared basis-state errors. Consequently,
\begin{equation}
    \label{eq:continuous_control_error}
    \left\|(V-T)P_{\mathcal E}\right\|\leq\left(\sum_{a=1}^{d}\left\|(V-T)\ket{e_a}\right\|^2\right)^{1/2}<\frac{\varepsilon}{2}.
\end{equation}

We next replace the continuous evolution by a finite PINEMAD sequence. On a segment of duration $\tau>0$ with constant control $g$, the Lie--Trotter product formula~\cite{trotterProductSemigroups1959} gives
\begin{equation}
    \label{eq:trotter}
    \begin{split}
        &\exp\left[\tau(-\mathrm{i}N^2+gb^\dagger-g^*b)\right]\\
        &\quad=\operatorname{s-lim}_{q\rightarrow\infty}\left[\exp\left(-\frac{\mathrm{i}\tau}{q}N^2\right)\exp\left(\frac{\tau}{q}(gb^\dagger-g^*b)\right)\right]^q\\
        &\quad=\operatorname{s-lim}_{q\rightarrow\infty}\left[U\left(\frac{\tau}{q}\right)A\left(\frac{\tau g}{q}\right)\right]^q.
    \end{split}
\end{equation}
The formula applies because $b$ is unitary and hence the coupling $gb^\dagger-g^*b$ is bounded. Each factor on the last line is one free drift following one PINEM interaction. The strong limit $\operatorname{s-lim}$ means convergence in norm on each fixed input state. There are finitely many constant-control segments, and all their propagators are unitary, so concatenating their product approximations also converges to $V$ on each fixed state. Since the basis of $\mathcal E$ is finite, one can choose finite splitting orders so that the same finite sequence $W_\ell$ has error below $\varepsilon/(2\sqrt d)$ on every basis state. The same finite-basis estimate then gives
\begin{equation}
    \label{eq:finite_sequence_error}
    \left\|(W_\ell-V)P_{\mathcal E}\right\|\leq\left(\sum_{a=1}^{d}\left\|(W_\ell-V)\ket{e_a}\right\|^2\right)^{1/2}<\frac{\varepsilon}{2}.
\end{equation}
Combining the continuous-control and splitting errors proves
\begin{equation}
    \left\|(W_\ell-T)P_{\mathcal E}\right\|\leq\left\|(W_\ell-V)P_{\mathcal E}\right\|+\left\|(V-T)P_{\mathcal E}\right\|<\varepsilon.
\end{equation}
Thus Theorem~1 follows if the BCS hypotheses hold for Eq.~(\ref{eq:auxiliary}). We now establish this: Sec.~\hyperref[sec:bcs_operators]{II.B} verifies the three basic operator assumptions, and Sec.~\hyperref[sec:bcs_lie]{II.C} verifies the Lie--Galerkin Control Condition.

\subsection{Verification of the basic operator hypotheses}
\label{sec:bcs_operators}

We first rewrite the three basic BCS hypotheses from Sec.~\hyperref[sec:bcs_result]{II.A} using our notation for Eq.~(\ref{eq:auxiliary}). The requirements are:
\begin{enumerate}
    \item \textit{(A1).} The drift generator $-\mathrm{i}N^2$ must have discrete spectrum with infinitely many distinct eigenvalues and a complete orthonormal eigenbasis $\{\ket n\}_{n\in\mathbb Z}$; degeneracies are allowed.
    \item \textit{(A2).} Every sideband state $\ket n$ must belong to the domains of both control generators $b^\dagger-b$ and $\mathrm{i}(b^\dagger+b)$.
    \item \textit{(A3).} For every constant $g$ with $\operatorname{Re}g,\operatorname{Im}g\in[-1,1]$, the generator $-\mathrm{i}N^2+gb^\dagger-g^*b$ must be essentially anti-Hermitian on
\begin{equation}
    \label{eq:pinemad_finite_span}
    \mathcal L=\left\{\sum_{n\in\mathbb Z}c_n\ket n:c_n\neq0\text{ for only finitely many }n\right\}\subsetneq\mathcal H.
\end{equation}
In physical terms, \textit{essentially anti-Hermitian} means that once the action of this generator on all finite sideband superpositions in $\mathcal L$ is specified, the unitary time evolution of every state in $\mathcal H$ is uniquely determined.
\end{enumerate}

We now verify these three requirements in the same order:
\begin{enumerate}
    \item \textit{Verification of (A1).} The states $\ket n$ diagonalize the drift $-\mathrm{i}N^2$, with eigenvalues $-\mathrm{i}n^2$. The degeneracy between $n$ and $-n$ is allowed.
    \item \textit{Verification of (A2).} Since $b$ is unitary, the two control generators $b^\dagger-b$ and $\mathrm{i}(b^\dagger+b)$ are bounded anti-Hermitian operators defined on all of $\mathcal H$.
    \item \textit{Verification of (A3).} We show that a unitary evolution exists and that the generator's action on finite sideband superpositions uniquely determines it. First, $N^2$ is self-adjoint on
\begin{equation}
    \mathcal D(N^2)=\left\{\sum_nc_n\ket n:\sum_nn^4|c_n|^2<\infty\right\}.
\end{equation}
    For any fixed allowed $g$, the coupling $gb^\dagger-g^*b$ is anti-Hermitian and bounded because $b$ and $b^\dagger$ are unitary:
\begin{equation}
    \label{eq:bcs_bounded_coupling}
    \left\|(gb^\dagger-g^*b)\ket\psi\right\|\leq2|g|\,\|\ket\psi\|,\qquad\ket\psi\in\mathcal H.
\end{equation}
    By the bounded-perturbation case of the Kato--Rellich theorem, the full generator $-\mathrm{i}N^2+gb^\dagger-g^*b$ is anti-Hermitian on $\mathcal D(N^2)$ and therefore generates a unitary time evolution on $\mathcal H$.

    We next show why specifying the generator on $\mathcal L$ is sufficient. For any $\ket\psi=\sum_nc_n\ket n\in\mathcal D(N^2)$, take the finite sideband approximation $\ket{\psi_M}=\sum_{|n|\leq M}c_n\ket n\in\mathcal L$. The state and the drift acting on it both converge in norm:
\begin{equation}
    \label{eq:bcs_truncation_limits}
    \begin{aligned}
        \|\ket\psi-\ket{\psi_M}\|^2&=\sum_{|n|>M}|c_n|^2\longrightarrow0,\\
        \|N^2(\ket\psi-\ket{\psi_M})\|^2&=\sum_{|n|>M}n^4|c_n|^2\longrightarrow0.
    \end{aligned}
\end{equation}
    The second limit follows from $\sum_nn^4|c_n|^2<\infty$. Since the coupling is bounded, the full generator acting on the truncated state also converges to its action on $\ket\psi$:
\begin{equation}
    \label{eq:bcs_full_generator_limit}
    \begin{split}
        &\left\|[-\mathrm{i}N^2+gb^\dagger-g^*b](\ket\psi-\ket{\psi_M})\right\|\\
        &\quad\leq\|N^2(\ket\psi-\ket{\psi_M})\|+2|g|\,\|\ket\psi-\ket{\psi_M}\|\longrightarrow0.
    \end{split}
\end{equation}
    These two convergences ensure that any anti-Hermitian generator agreeing with the prescribed action on $\mathcal L$ must also agree with it throughout $\mathcal D(N^2)$. Since the full generator on this domain is already anti-Hermitian, no different anti-Hermitian generator can extend that action, and the resulting unitary evolution is unique. This proves (A3).
\end{enumerate}

\subsection{Verification of the Lie--Galerkin condition}
\label{sec:bcs_lie}
\subsubsection{Definition of the Lie--Galerkin Control Condition in PINEMAD}

The general Lie--Galerkin Control Condition in BCS Definition~2.5~\cite{boscainMultiinputSchrodingerEquation2014} involves several technical definitions for arbitrary drift spectra and control ranges. Here we give only a sufficient formulation for PINEMAD.

We first establish the notation. For each positive integer cutoff $M$, define
\begin{equation}
    \mathcal H_M=\operatorname{span}\{\ket{n}:|n|\leq M\}.
\end{equation}
Order the drift eigenstates as $\ket0,\ket{-1},\ket1,\ket{-2},\ket2,\ldots$. The first $2M+1$ states then span $\mathcal H_M$, so these windows coincide with the finite-dimensional truncations used by BCS.

Let $P_M$ project onto this window. For the repeated matrix calculations below, denote the three projected generators by $\mathsf A_M$, $\mathsf B_{1,M}$, and $\mathsf B_{2,M}$:
\begin{equation}
    \begin{split}
        \mathsf A_M&=-\mathrm{i}P_MN^2P_M =-\mathrm{i}\sum_{n=-M}^{M}n^2\ket{n}\bra{n},\\
        \mathsf B_{1,M}&=P_M(b^\dagger-b)P_M =\sum_{n=-M}^{M-1}\left(\ket{n+1}\bra{n}-\ket{n}\bra{n+1}\right),\\
        \mathsf B_{2,M}&=\mathrm{i}P_M(b^\dagger+b)P_M =\mathrm{i}\sum_{n=-M}^{M-1}\left(\ket{n+1}\bra{n}+\ket{n}\bra{n+1}\right).
    \end{split}
\end{equation}
We also define
\begin{equation}
    \label{eq:pinemad_gap_component}
    \bra m\mathsf B_j^{[\sigma]}\ket n=\begin{cases}
        \bra m\mathsf B_j\ket n,&|m^2-n^2|=\sigma,\\
        0,&|m^2-n^2|\neq\sigma,
    \end{cases}
\end{equation}
Here $m,n\in\mathbb Z$ are sideband indices, $j=1,2$ labels the control generators $\mathsf B_1=b^\dagger-b$ and $\mathsf B_2=\mathrm{i}(b^\dagger+b)$, and $\sigma\geq0$ specifies the absolute difference $|m^2-n^2|$ between the quadratic drift-phase coefficients. Both $\mathsf B_j$ and $\mathsf B_j^{[\sigma]}$ act on the full sideband space $\mathcal H$.

For PINEMAD with the symmetric control ranges used here, the Lie--Galerkin Control Condition is ensured by the following two requirements for arbitrarily large $M$:
\begin{enumerate}
    \item For $j=1,2$ and $r=0,\ldots,M-1$, require
\begin{equation}
    \label{eq:pinemad_gap_no_leakage}
    (1-P_M)\mathsf B_j^{[2r+1]}P_M=0,\qquad P_M\mathsf B_j^{[2r+1]}(1-P_M)=0.
\end{equation}
    Here $1-P_M$ projects onto the sidebands outside the window. Also require
\begin{equation}
    \label{eq:pinemad_gap_reconstruction}
    \mathsf B_{j,M}=\sum_{r=0}^{M-1}P_M\mathsf B_j^{[2r+1]}P_M,\qquad j=1,2.
\end{equation}
    \item The three projected generators must satisfy
\begin{equation}
    \label{eq:pinemad_lg_requirement}
    \operatorname{Lie}_{\mathbb R}\{\mathsf A_M,\mathsf B_{1,M},\mathsf B_{2,M}\}\supseteq\mathfrak{su}(2M+1).
\end{equation}
\end{enumerate}
Here $\operatorname{Lie}_{\mathbb R}$ denotes the real linear combinations and repeated commutators generated by these matrices, and $\mathfrak{su}(2M+1)$ is the space of traceless anti-Hermitian matrices on $\mathcal H_M$.

We now verify these two requirements:

\subsubsection{Verification of requirement 1 of the Lie--Galerkin condition}

We verify Eqs.~(\ref{eq:pinemad_gap_no_leakage}) and~(\ref{eq:pinemad_gap_reconstruction}) directly for $\mathsf B_j^{[2r+1]}$, with $r=0,\ldots,M-1$. By Eq.~(\ref{eq:pinemad_gap_component}), these operators retain the matrix elements with $|m^2-n^2|=2r+1$. Since $\mathsf B_1$ and $\mathsf B_2$ couple only neighboring sidebands, the contributing pairs $(n,n+1)$ are determined by
\begin{equation}
    \label{eq:pinemad_selected_pairs}
    |(n+1)^2-n^2|=|2n+1|=2r+1\quad\Longleftrightarrow\quad n=r\ \text{or}\ n=-r-1.
\end{equation}
Thus the full-space operators defined in Eq.~(\ref{eq:pinemad_gap_component}) are
\begin{equation}
    \label{eq:pinemad_selected_components}
    \begin{aligned}
        \mathsf B_1^{[2r+1]}&=\sum_{n\in\{r,-r-1\}}\left(\ket{n+1}\bra n-\ket n\bra{n+1}\right),\\
        \mathsf B_2^{[2r+1]}&=\mathrm{i}\sum_{n\in\{r,-r-1\}}\left(\ket{n+1}\bra n+\ket n\bra{n+1}\right).
    \end{aligned}
\end{equation}
For $r=0,\ldots,M-1$, all four sideband indices $r,r+1,-r-1,-r$ lie in $[-M,M]$. Hence each selected operator acts entirely within $\mathcal H_M$:
\begin{equation}
    \label{eq:pinemad_selected_support}
    \mathsf B_j^{[2r+1]}=P_M\mathsf B_j^{[2r+1]}P_M,\qquad j=1,2.
\end{equation}
Multiplication by $1-P_M$ on either side gives the two zero blocks in Eq.~(\ref{eq:pinemad_gap_no_leakage}).

As $r$ runs from $0$ to $M-1$, the indices $r$ and $-r-1$ cover every $n=-M,\ldots,M-1$ exactly once. Summing the selected operators therefore includes every neighboring-sideband coupling inside the window exactly once:
\begin{equation}
    \label{eq:pinemad_reconstruction_verified}
    \sum_{r=0}^{M-1}P_M\mathsf B_j^{[2r+1]}P_M=P_M\mathsf B_jP_M=\mathsf B_{j,M},\qquad j=1,2.
\end{equation}
This proves Eq.~(\ref{eq:pinemad_gap_reconstruction}) and completes the verification of requirement 1.

\subsubsection{Verification of requirement 2 of the Lie--Galerkin condition}

We verify Eq.~(\ref{eq:pinemad_lg_requirement}) by first separating the couplings between individual neighboring sidebands, then generating relative phases and couplings between more distant sidebands. For $-M\leq n<M$, define
\begin{equation}
    F_n=\ket{n+1}\bra n-\ket n\bra{n+1},\qquad G_n=\mathrm{i}\left(\ket{n+1}\bra n+\ket n\bra{n+1}\right).
\end{equation}
These operators represent the two phase quadratures of the coupling between sidebands $n$ and $n+1$, and the projected controls are
\begin{equation}
    \mathsf B_{1,M}=\sum_{n=-M}^{M-1}F_n,\qquad \mathsf B_{2,M}=\sum_{n=-M}^{M-1}G_n.
\end{equation}
Direct commutation with the drift gives
\begin{equation}
    [\mathsf A_M,F_n]=-(2n+1)G_n,\qquad [\mathsf A_M,G_n]=(2n+1)F_n.
\end{equation}
Each commutation adds one power of the signed spacing $2n+1$. More explicitly, for every integer $m\geq0$,
\begin{equation}
    \label{eq:pinemad_repeated_commutator}
    \begin{aligned}
        \left[\mathsf A_M,\sum_{n=-M}^{M-1}(2n+1)^mF_n\right]&=-\sum_{n=-M}^{M-1}(2n+1)^{m+1}G_n,\\
        \left[\mathsf A_M,\sum_{n=-M}^{M-1}(2n+1)^mG_n\right]&=\sum_{n=-M}^{M-1}(2n+1)^{m+1}F_n.
    \end{aligned}
\end{equation}
The $m=0$ sums are $\mathsf B_{1,M}$ and $\mathsf B_{2,M}$. Applying the two commutator identities recursively gives, for $m=0,\ldots,2M-1$,
\begin{equation}
    \label{eq:pinemad_weighted_membership}
    \begin{aligned}
        \sum_{n=-M}^{M-1}(2n+1)^mF_n&\in\operatorname{Lie}_{\mathbb R}\{\mathsf A_M,\mathsf B_{1,M},\mathsf B_{2,M}\},\\
        \sum_{n=-M}^{M-1}(2n+1)^mG_n&\in\operatorname{Lie}_{\mathbb R}\{\mathsf A_M,\mathsf B_{1,M},\mathsf B_{2,M}\}.
    \end{aligned}
\end{equation}
For either $F_n$ or $G_n$, these are $2M$ linear equations for the $2M$ individual coupling generators. Explicitly,
\begin{equation}
    \label{eq:pinemad_vandermonde_F}
    \begin{pmatrix}
        1&1&\cdots&1\\
        -2M+1&-2M+3&\cdots&2M-1\\
        \vdots&\vdots&\ddots&\vdots\\
        (-2M+1)^{2M-1}&(-2M+3)^{2M-1}&\cdots&(2M-1)^{2M-1}
    \end{pmatrix}
    \begin{pmatrix}
        F_{-M}\\
        F_{-M+1}\\
        \vdots\\
        F_{M-1}
    \end{pmatrix}
    =\begin{pmatrix}
        \sum_{n=-M}^{M-1}F_n\\
        \sum_{n=-M}^{M-1}(2n+1)F_n\\
        \vdots\\
        \sum_{n=-M}^{M-1}(2n+1)^{2M-1}F_n
    \end{pmatrix}.
\end{equation}
\begin{equation}
    \label{eq:pinemad_vandermonde_G}
    \begin{pmatrix}
        1&1&\cdots&1\\
        -2M+1&-2M+3&\cdots&2M-1\\
        \vdots&\vdots&\ddots&\vdots\\
        (-2M+1)^{2M-1}&(-2M+3)^{2M-1}&\cdots&(2M-1)^{2M-1}
    \end{pmatrix}
    \begin{pmatrix}
        G_{-M}\\
        G_{-M+1}\\
        \vdots\\
        G_{M-1}
    \end{pmatrix}
    =\begin{pmatrix}
        \sum_{n=-M}^{M-1}G_n\\
        \sum_{n=-M}^{M-1}(2n+1)G_n\\
        \vdots\\
        \sum_{n=-M}^{M-1}(2n+1)^{2M-1}G_n
    \end{pmatrix}.
\end{equation}
The common coefficient matrix has entries $(2n+1)^m$, with $m=0,\ldots,2M-1$ and $n=-M,\ldots,M-1$, and its Vandermonde determinant is
\begin{equation}
    \prod_{-M\leq n<k\leq M-1}\left[(2k+1)-(2n+1)\right]\neq0,
\end{equation}
because all signed spacings $2n+1$ are different. Inverting this coefficient matrix expresses every individual $F_n$ and $G_n$ as a real linear combination of the preceding sums. Therefore,
\begin{equation}
    \label{eq:pinemad_individual_membership}
    F_n,G_n\in\operatorname{Lie}_{\mathbb R}\{\mathsf A_M,\mathsf B_{1,M},\mathsf B_{2,M}\},\qquad -M\leq n<M.
\end{equation}

The commutator
\begin{equation}
    [F_n,G_n]=2\mathrm{i}\left(\ket{n+1}\bra{n+1}-\ket{n}\bra{n}\right)
\end{equation}
provides the relative-phase generator for adjacent sidebands. Hence,
\begin{equation}
    \label{eq:pinemad_phase_membership}
    \mathrm{i}\left(\ket{n+1}\bra{n+1}-\ket n\bra n\right)\in\operatorname{Lie}_{\mathbb R}\{\mathsf A_M,\mathsf B_{1,M},\mathsf B_{2,M}\},\qquad -M\leq n<M.
\end{equation}
Real linear combinations of these adjacent phase differences provide every relative phase within the window. Couplings between more distant sidebands follow by joining neighboring transitions. For $-M\leq n<M-1$,
\begin{equation}
    \begin{aligned}
        [F_n,F_{n+1}]&=-\left(\ket{n+2}\bra n-\ket n\bra{n+2}\right),\\
        [F_n,G_{n+1}]&=-\mathrm{i}\left(\ket{n+2}\bra n+\ket n\bra{n+2}\right).
    \end{aligned}
\end{equation}
For these next-neighbor couplings, it follows that
\begin{equation}
    \label{eq:pinemad_next_neighbor_membership}
    \begin{aligned}
        \ket{n+2}\bra n-\ket n\bra{n+2}&\in\operatorname{Lie}_{\mathbb R}\{\mathsf A_M,\mathsf B_{1,M},\mathsf B_{2,M}\},\\
        \mathrm{i}\left(\ket{n+2}\bra n+\ket n\bra{n+2}\right)&\in\operatorname{Lie}_{\mathbb R}\{\mathsf A_M,\mathsf B_{1,M},\mathsf B_{2,M}\}.
    \end{aligned}
\end{equation}
Commuting them with the coupling to the next sideband extends the transition by one more step. Repeating along the ladder gives, for every $-M\leq n<m\leq M$,
\begin{equation}
    \label{eq:pinemad_all_pair_membership}
    \begin{aligned}
        \ket m\bra n-\ket n\bra m&\in\operatorname{Lie}_{\mathbb R}\{\mathsf A_M,\mathsf B_{1,M},\mathsf B_{2,M}\},\\
        \mathrm{i}\left(\ket m\bra n+\ket n\bra m\right)&\in\operatorname{Lie}_{\mathbb R}\{\mathsf A_M,\mathsf B_{1,M},\mathsf B_{2,M}\}.
    \end{aligned}
\end{equation}
Together with the diagonal generators for relative phases, these coupling operators span $\mathfrak{su}(2M+1)$. Hence,
\begin{equation}
    \label{eq:pinemad_lg_verified}
    \operatorname{Lie}_{\mathbb R}\{\mathsf A_M,\mathsf B_{1,M},\mathsf B_{2,M}\}\supseteq\mathfrak{su}(2M+1).
\end{equation}
This is Eq.~(\ref{eq:pinemad_lg_requirement}), so requirement 2 is satisfied.

Both requirements hold for arbitrarily large $M$. Together with the operator hypotheses verified in Sec.~\hyperref[sec:bcs_operators]{II.B}, this establishes all hypotheses of BCS Theorem~2.6. The argument in Sec.~\hyperref[sec:bcs_result]{II.A} therefore proves Theorem~1.

\section{Two-sideband state preparation}
\label{sec:two_sideband}

To illustrate state preparation with a small number of PINEMAD stages, we consider the ideal-sideband input $\ket0$ and the target
\begin{equation}
    \label{eq:two_sideband_target}
    \ket{\psi_{\mathrm{tar}}(\varphi)}=\frac{\ket0+\exp(\ii\varphi)\ket1}{\sqrt2},\qquad 0\leq\varphi\leq2\pi.
\end{equation}
We optimize the state fidelity $\mathcal F=|\bra{\psi_{\mathrm{tar}}(\varphi)}W_\ell\ket0|^2$ under $|g_j|\leq2$ and $0\leq\theta_j\leq2\pi$. The sequences with $\ell=2$ and $3$ achieve $\mathcal F=0.962956027$ and $0.998979831$, respectively.

The achievable fidelities are independent of the target relative phase $\varphi$. This follows from the transformation properties of the PINEM and drift operators:
\begin{equation}
    \label{eq:two_sideband_covariance}
    \exp(\ii\alpha N)A(g)\exp(-\ii\alpha N)=A(g\exp(\ii\alpha)),\qquad [U(\theta),\exp(\ii\alpha N)]=0.
\end{equation}
Since $\exp(\ii\alpha N)\ket0=\ket0$, multiplying every coupling $g_j$ by $\exp(\ii\alpha)$ multiplies the output coefficient of sideband $n$ by $\exp(\ii n\alpha)$. The target transforms in the same way:
\begin{equation}
    \label{eq:two_sideband_target_phase}
    \exp(\ii\alpha N)\ket{\psi_{\mathrm{tar}}(\varphi)}=\ket{\psi_{\mathrm{tar}}(\varphi+\alpha)}.
\end{equation}
The state overlap is therefore unchanged. Starting from a sequence optimized for $\varphi=0$, any target phase $\varphi$ is obtained by replacing every $g_j$ with $g_j\exp(\ii\varphi)$ while retaining the same drift phases. This common phase adjustment preserves the coupling bounds, so the same fidelities are achieved for every target relative phase.

\section{Finite-energy-spread dynamics and fidelity}\label{sec:finite}

\subsection{Energy normalization and shifted wavepackets}

For the continuum model, take energy-normalized states satisfying $\braket{E'|E}=\delta(E'-E)$, where $\delta$ denotes the Dirac delta distribution. In the main text, we use the notation
\begin{equation}
    \label{eq:energy_normalization}
    x=\frac{E-E_0}{\hbar\omega_0},\qquad \ket{x}:=\ket{E_0+x\hbar\omega_0},\qquad \braket{y|x}=\frac{\delta(y-x)}{\hbar\omega_0}.
\end{equation}
These $\ket{x}$ retain the energy normalization, without an additional rescaling.

For an unchirped incident electron, the normalized Gaussian amplitude is chosen real:
\begin{equation}
    f(\epsilon,\sigma_E) =\frac{\exp[-\epsilon^2/(4\sigma_E^2)]} {(2\pi\sigma_E^2)^{1/4}}, \qquad \ket{\Psi_{\mathrm{in},\sigma_E}} =\int_{-\infty}^{+\infty}\mathrm{d}E\, f(E-E_0,\sigma_E)\ket{E}.
\end{equation}
The incident wavepacket is centered at the reference kinetic energy $E_0$, and $\sigma_E$ is the standard deviation of its energy probability density $|f|^2$. The FWHM of this distribution is $\Delta E=2\sqrt{2\ln2}\,\sigma_E$. Define the Gaussian centered on sideband $n$ by
\begin{equation}
    \ket{\Phi_{n,\sigma_E}} =\int_{-\infty}^{+\infty}\mathrm{d}E\, f(E-E_0-n\hbar\omega_0,\sigma_E)\ket{E}.
\end{equation}
For an ideal state $\ket{\psi}=\sum_nc_n\ket{n}$, introduce the linear Gaussian-envelope map $Q_{\sigma_E}$:
\begin{equation}
    \label{eq:gaussian_map}
    Q_{\sigma_E}\ket{\psi} =\sum_nc_n\ket{\Phi_{n,\sigma_E}}.
\end{equation}
The sum is interpreted in the continuum Hilbert-space norm. Boundedness for infinitely supported coefficients is established in Sec.~\hyperref[sec:corollary]{V}. The overlaps are
\begin{equation}
    \label{eq:gaussian_gram}
    \braket{\Phi_{m,\sigma_E}|\Phi_{n,\sigma_E}} =\exp\left[-\frac{(m-n)^2(\hbar\omega_0)^2} {8\sigma_E^2}\right].
\end{equation}
Thus the normalized finite-width counterpart is
\begin{equation}
    \ket{\Psi_{\psi,\sigma_E}} =\frac{Q_{\sigma_E}\ket{\psi}} {\|Q_{\sigma_E}\ket{\psi}\|}.
\end{equation}

For the ideal target $\ket{\psi_{\mathrm{tar}}}=\sum_nc_n\ket{n}$, the main text uses
\begin{equation}
    \ket{\Psi_{\mathrm{tar},\sigma_E}} =\mathcal N\int_{-\infty}^{+\infty}\mathrm{d}x\, f(x\hbar\omega_0,\sigma_E)\sum_nc_n\ket{x+n}.
\end{equation}
Since $\mathrm{d}E=\hbar\omega_0\,\mathrm{d}x$, the normalization constant in this convention is
\begin{equation}
    \label{eq:target_normalization}
    \mathcal N =\hbar\omega_0 \left\{\sum_{m,n}c_m^*c_n \exp\left[-\frac{(m-n)^2(\hbar\omega_0)^2} {8\sigma_E^2}\right]\right\}^{-1/2}.
\end{equation}
Although its subscript is omitted, $\mathcal N$ depends on $\sigma_E$ and the target coefficients. The construction is a coherent sum of amplitudes, with overlapping shifted envelopes retained.

\subsection{Action of a PINEMAD sequence}

Here we derive the action of a PINEMAD sequence on an electron wavepacket with finite energy spread.

The PINEM coupling is assumed approximately energy independent across the populated spectrum. The energy translation operator $b$ acts as $b\ket{x}=\ket{x-1}$, while $U_x(\theta)\ket{x}=\exp(-\mathrm{i}\theta x^2)\ket{x}$. Fix an incident offset $x$. On the translated components $\ket{x+m}$, identify the integer $m$ with an auxiliary discrete ladder. Its drift is
\begin{equation}
    \label{eq:offset_drift}
    U_x(\theta)=\exp[-\mathrm{i}\theta(N+x)^2] =\exp(-\mathrm{i}\theta x^2)U(\theta)R_x(\theta), \qquad R_x(\theta)=\exp(-2\mathrm{i}\theta xN).
\end{equation}
The rotations $R_x$ commute with the ideal drift and obey
\begin{equation}
    \label{eq:rotation}
    R_x(\theta)A(g) =A\left(g\exp[-2\mathrm{i}\theta x]\right)R_x(\theta).
\end{equation}
This identity follows from $R_x(\theta)b^\dagger R_x(\theta)^\dagger =\exp(-2\mathrm{i}\theta x)b^\dagger$ and its adjoint.

For a given sequence, define
\begin{equation}
    \Theta=\sum_{j=1}^{\ell}\theta_j,\qquad \widetilde g_j(x) =g_j\exp\left[-2\mathrm{i}x\sum_{k=j}^{\ell}\theta_k\right].
\end{equation}
Using Eq.~(\ref{eq:rotation}), we move all rotations to the right. The rotation from stage $k$ passes through every PINEM operator with index $j\leq k$, giving the accumulated phase $\sum_{k=j}^{\ell}\theta_k$ in $\widetilde g_j(x)$. Thus,
\begin{equation}
    \label{eq:offset_sequence}
    U_x(\theta_\ell)A(g_\ell)\cdots U_x(\theta_1)A(g_1) =\exp(-\mathrm{i}\Theta x^2)\mathcal W(x)R_x(\Theta),
\end{equation}
where the auxiliary discrete-ladder operator is
\begin{equation}
    \mathcal W(x) =U(\theta_\ell)A[\widetilde g_\ell(x)]\cdots U(\theta_1)A[\widetilde g_1(x)].
\end{equation}
The modified couplings describe this rearranged representation; the physical couplings remain $g_j$. The incident component $\ket{x}$ corresponds to the auxiliary input $\ket{0}$, for which $R_x(\Theta)\ket{0}=\ket{0}$. Denote the physical sequence acting on the continuous energy spectrum by $W_\ell^{\mathrm{cont}}$. With $\mathcal W_{n0}(x)=\bra{n}\mathcal W(x)\ket{0}$ calculated in the ideal discrete-sideband basis,
\begin{equation}
    W_\ell^{\mathrm{cont}}\ket{x} =\exp(-\mathrm{i}\Theta x^2) \sum_n\mathcal W_{n0}(x)\ket{x+n}.
\end{equation}
Integrating this action over the incident Gaussian wavepacket and using $\mathrm{d}E=\hbar\omega_0\,\mathrm{d}x$ yields
\begin{equation}
    \label{eq:output}
    \begin{aligned}
        \ket{\Psi_{\mathrm{out},\sigma_E}} &=\int_{-\infty}^{+\infty}\mathrm{d}x\, F(x,\sigma_E)\sum_n\mathcal W_{n0}(x)\ket{x+n},\\
        F(x,\sigma_E) &=\hbar\omega_0f(x\hbar\omega_0,\sigma_E) \exp(-\mathrm{i}\Theta x^2).
    \end{aligned}
\end{equation}
The phase $\exp(-\mathrm{i}\Theta x^2)$ is part of the output amplitude and is retained when comparing against the unchirped target.

\subsection{Full-wavepacket fidelity}

Write
\begin{equation}
    \ket{\Psi_{\mathrm{tar},\sigma_E}} =\int\mathrm{d}E\, \psi_{\mathrm{tar},\sigma_E}(E)\ket{E}, \qquad \ket{\Psi_{\mathrm{out},\sigma_E}} =\int\mathrm{d}E\, \psi_{\mathrm{out},\sigma_E}(E)\ket{E}.
\end{equation}
where the scalar energy amplitudes are
\begin{equation}
    \psi_{\mathrm{tar},\sigma_E}(E) =\frac{\mathcal N}{\hbar\omega_0} \sum_mc_mf(E-E_0-m\hbar\omega_0,\sigma_E)
\end{equation}
and
\begin{equation}
    \psi_{\mathrm{out},\sigma_E}(E) =\sum_n f(E-E_0-n\hbar\omega_0,\sigma_E) \exp\left[-\mathrm{i}\Theta \left(\frac{E-E_0}{\hbar\omega_0}-n\right)^2\right] \mathcal W_{n0}\left( \frac{E-E_0}{\hbar\omega_0}-n\right).
\end{equation}
The factor $\mathcal N/(\hbar\omega_0)$ follows from the main-text target definition, in which $\mathcal N$ absorbs the Jacobian between the energy and dimensionless-energy integrals.

To quantify the agreement between the normalized output and target wavepackets, we define the state-preparation fidelity
\begin{equation}
    \label{eq:full_fidelity}
    \mathcal F_{\sigma_E} =\left|\int_{-\infty}^{+\infty}\mathrm{d}E\, \psi_{\mathrm{tar},\sigma_E}(E)^* \psi_{\mathrm{out},\sigma_E}(E)\right|^2.
\end{equation}
Equivalently,
\begin{equation}
    \label{eq:fidelity_x}
    \mathcal F_{\sigma_E} =\frac{\mathcal N^2}{(\hbar\omega_0)^2} \left|\sum_{m,n}c_m^* \int_{-\infty}^{+\infty}\mathrm{d}x\, f[(x+n-m)\hbar\omega_0,\sigma_E]^* F(x,\sigma_E)\mathcal W_{n0}(x)\right|^2.
\end{equation}
This fidelity retains coherent interference between overlapping shifted envelopes. It therefore differs from an incoherent average of discrete-ladder fidelities over incident energies.

\section{Proof of Corollary 1}\label{sec:corollary}

Let $\mathcal E$ and $T$ satisfy Theorem~1. For every $\varepsilon>0$, we seek one finite PINEMAD sequence and a width threshold $\sigma_E^*>0$ such that
\begin{equation}
    \label{eq:corollary_statement}
    \left|\braket{\Psi_{T\psi,\sigma_E}|W_\ell^{\mathrm{cont}}|\Psi_{\psi,\sigma_E}}\right|^2>1-\varepsilon
\end{equation}
for every normalized $\ket\psi\in\mathcal E$ and every $0<\sigma_E<\sigma_E^*$. The sequence must be the same for all these inputs and widths.

We first recall the notation of Sec.~\hyperref[sec:finite]{IV}. The map $Q_{\sigma_E}$ replaces each ideal sideband by a shifted Gaussian wavepacket with envelope $f$, and the normalized input and target are
\begin{equation}
    \label{eq:finite_width_counterparts}
    \ket{\Psi_{\psi,\sigma_E}}=\frac{Q_{\sigma_E}\ket\psi}{\|Q_{\sigma_E}\ket\psi\|},\qquad\ket{\Psi_{T\psi,\sigma_E}}=\frac{Q_{\sigma_E}T\ket\psi}{\|Q_{\sigma_E}T\ket\psi\|}.
\end{equation}
The ideal sequence is $W_\ell$, while $W_\ell^{\mathrm{cont}}$ denotes the same physical sequence acting on the continuous energy spectrum. We compare the actual finite-width output $W_\ell^{\mathrm{cont}}Q_{\sigma_E}\ket\psi$ with $Q_{\sigma_E}W_\ell\ket\psi$, obtained by broadening the ideal output. For any fixed sequence, we will prove
\begin{equation}
    \label{eq:corollary_key_limits}
    \begin{aligned}
        \left\|Q_{\sigma_E}^\dagger Q_{\sigma_E}-\mathds1\right\|&\longrightarrow0,\\
        \left\|\left(W_\ell^{\mathrm{cont}}Q_{\sigma_E}-Q_{\sigma_E}W_\ell\right)P_{\mathcal E}\right\|&\longrightarrow0
    \end{aligned}
\end{equation}
as $\sigma_E\to0$. The first limit controls normalization; the second controls the evolution error for all normalized inputs in $\mathcal E$. The proof proceeds in three parts:
\begin{enumerate}
    \item \textit{Normalization, Sec. IV.A.} Use the overlaps of shifted Gaussians to bound the change in norm, including for states with infinitely many populated sidebands.
    \item \textit{Evolution, Sec. IV.B.} Write the wavepacket as sideband ladders with fractional energy offsets. Bound the Gaussian tails and compare the fixed-sequence evolution near zero offset.
    \item \textit{Fidelity, Sec. IV.C.} Choose the ideal sequence using Theorem~1, then reduce the width until the evolution and normalization errors give the required fidelity.
\end{enumerate}

\subsection{Normalization after Gaussian broadening}
\label{sec:width_normalization}

We prove the first limit in Eq.~(\ref{eq:corollary_key_limits}). The ideal sidebands are orthogonal, but their Gaussian counterparts overlap. Equation~(\ref{eq:gaussian_gram}) gives
\begin{equation}
    \bra mQ_{\sigma_E}^\dagger Q_{\sigma_E}\ket n=\exp\left[-\frac{(m-n)^2(\hbar\omega_0)^2}{8\sigma_E^2}\right].
\end{equation}
For a finite sideband superposition $\ket\psi=\sum_nc_n\ket n$, the diagonal terms give $\|\ket\psi\|^2$, and the remaining terms satisfy
\begin{equation}
    \label{eq:gaussian_norm_estimate}
    \begin{aligned}
        \left|\|Q_{\sigma_E}\ket\psi\|^2-\|\ket\psi\|^2\right|&\leq\sum_{k\neq0}\exp\left[-\frac{k^2(\hbar\omega_0)^2}{8\sigma_E^2}\right]\left|\sum_nc_{n+k}^*c_n\right|\\
        &\leq2\sum_{k=1}^{\infty}\exp\left[-\frac{k^2(\hbar\omega_0)^2}{8\sigma_E^2}\right]\|\ket\psi\|^2.
    \end{aligned}
\end{equation}
The last step uses Cauchy--Schwarz:
\begin{equation}
    \left|\sum_nc_{n+k}^*c_n\right|\leq\left(\sum_n|c_{n+k}|^2\right)^{1/2}\left(\sum_n|c_n|^2\right)^{1/2}=\|\ket\psi\|^2.
\end{equation}
Thus the bound is independent of how many sidebands are populated.

The same estimate also makes the Gaussian construction well defined for an arbitrary $\ket\psi\in\mathcal H$. For two finite cutoffs $M'>M$, it bounds the change caused by adding more sidebands:
\begin{equation}
    \label{eq:gaussian_cutoff_convergence}
    \left\|Q_{\sigma_E}\sum_{M<|n|\leq M'}c_n\ket n\right\|^2\leq\left(1+2\sum_{k=1}^{\infty}\exp\left[-\frac{k^2(\hbar\omega_0)^2}{8\sigma_E^2}\right]\right)\sum_{|n|>M}|c_n|^2\longrightarrow0.
\end{equation}
Here $\sigma_E$ is fixed and $M\to\infty$. Since $\sum_n|c_n|^2<\infty$, the right-hand side tends to zero independently of $M'$. The broadened finite sums therefore converge in norm, and Eq.~(\ref{eq:gaussian_norm_estimate}) holds for every state in $\mathcal H$.

For a bounded Hermitian operator, its norm can be evaluated from the expectation values on normalized states. Consequently,
\begin{equation}
    \label{eq:gram_bound}
    \begin{aligned}
        \left\|Q_{\sigma_E}^\dagger Q_{\sigma_E}-\mathds1\right\|&=\sup_{\|\ket\psi\|=1}\left|\|Q_{\sigma_E}\ket\psi\|^2-1\right|\\
        &\leq2\sum_{k=1}^{\infty}\exp\left[-\frac{k^2(\hbar\omega_0)^2}{8\sigma_E^2}\right].
    \end{aligned}
\end{equation}
Using $k^2\geq k$ for $k\geq1$ gives an explicit width-dependent bound:
\begin{equation}
    2\sum_{k=1}^{\infty}\exp\left[-\frac{k^2(\hbar\omega_0)^2}{8\sigma_E^2}\right]\leq\frac{2}{\exp[(\hbar\omega_0)^2/(8\sigma_E^2)]-1}\longrightarrow0.
\end{equation}
In particular,
\begin{equation}
    \label{eq:norm_bound}
    \left|\|Q_{\sigma_E}\ket\psi\|^2-1\right|\leq\left\|Q_{\sigma_E}^\dagger Q_{\sigma_E}-\mathds1\right\|\longrightarrow0,\qquad\|\ket\psi\|=1,
\end{equation}
and
\begin{equation}
    \label{eq:gaussian_operator_norm_limit}
    \left|\|Q_{\sigma_E}\|^2-1\right|\leq\left\|Q_{\sigma_E}^\dagger Q_{\sigma_E}-\mathds1\right\|\longrightarrow0.
\end{equation}
These bounds apply to both the input $\ket\psi$ and the ideal target $T\ket\psi$, since $T$ preserves the ideal norm.

For sufficiently small width, every nonzero state also has a nonzero Gaussian-counterpart norm:
\begin{equation}
    \|Q_{\sigma_E}\ket\psi\|^2\geq\left(1-\|Q_{\sigma_E}^\dagger Q_{\sigma_E}-\mathds1\|\right)\|\ket\psi\|^2>0.
\end{equation}
A sufficient condition follows from the geometric-series bound above:
\begin{equation}
    \label{eq:normalizability_condition}
    \frac{\sigma_E}{\hbar\omega_0}<\frac{1}{\sqrt{8\ln3}},\qquad\text{equivalently}\qquad\Delta E<\hbar\omega_0\sqrt{\frac{\ln2}{\ln3}}\approx0.79\hbar\omega_0.
\end{equation}
This guarantees that the normalized states in Eq.~(\ref{eq:finite_width_counterparts}) are defined. The fidelity requirement will determine how much smaller the width needs to be.

\subsection{Comparison of continuum and ideal dynamics}
\label{sec:width_dynamics}

We now prove the second limit in Eq.~(\ref{eq:corollary_key_limits}), keeping every $g_j$ and $\theta_j$ fixed. The proof separates the Gaussian tail error from the change in evolution near the central energy.

\subsubsection{Energy offsets and the Gaussian wavepacket}

Using the dimensionless energy offset of Sec.~\hyperref[sec:finite]{IV}, write each energy uniquely as
\begin{equation}
    \label{eq:fractional_energy_offset}
    E=E_0+(n+x)\hbar\omega_0,\qquad n\in\mathbb Z,\quad-\frac12\leq x<\frac12.
\end{equation}
Here $x$ is restricted to one energy interval, and the integer $n$ specifies the sideband. PINEM changes $n$ by integers, while drift changes phases, so neither operation mixes different values of this fractional offset. For any continuum wavepacket $\ket\Psi$, energy normalization gives
\begin{equation}
    \|\ket\Psi\|^2=\hbar\omega_0\int_{-1/2}^{1/2}\mathrm dx\sum_n\left|\braket{E_0+(n+x)\hbar\omega_0|\Psi}\right|^2.
\end{equation}
On one offset ladder, the amplitudes of $Q_{\sigma_E}\ket\psi$ can be written in the ideal discrete basis:
\begin{equation}
    \label{eq:Q_fibers}
    \sum_n\ket n\bra{E_0+(n+x)\hbar\omega_0}Q_{\sigma_E}\ket\psi=\sum_{k\in\mathbb Z}f((x+k)\hbar\omega_0,\sigma_E)(b^\dagger)^k\ket\psi.
\end{equation}
To obtain this expression, insert the shifted Gaussians of Sec.~\hyperref[sec:finite]{IV} and set $k=n-m$. Negative powers of the unitary $b^\dagger$ represent downward sideband shifts. The $k=0$ term is $f(x\hbar\omega_0,\sigma_E)\ket\psi$; the terms with $k\neq0$ are tails of Gaussians centered in other energy intervals.

\subsubsection{Gaussian tails outside the central interval}

For every normalized $\ket\psi$, the triangle inequality and the norm-preserving sideband shifts give
\begin{equation}
    \label{eq:tail_bound}
    \begin{split}
        &\left[\hbar\omega_0\int_{-1/2}^{1/2}\mathrm dx\left\|\sum_{k\neq0}f((x+k)\hbar\omega_0,\sigma_E)(b^\dagger)^k\ket\psi\right\|^2\right]^{1/2}\\
        &\quad\leq\sum_{k\neq0}\left[\hbar\omega_0\int_{-1/2}^{1/2}\mathrm dx\left|f((x+k)\hbar\omega_0,\sigma_E)\right|^2\right]^{1/2}\\
        &\quad\leq2\sqrt{\hbar\omega_0}(2\pi\sigma_E^2)^{-1/4}\frac{\exp[-(\hbar\omega_0)^2/(16\sigma_E^2)]}{1-\exp[-(\hbar\omega_0)^2/(16\sigma_E^2)]}\longrightarrow0.
    \end{split}
\end{equation}
The last step uses $|x+k|\geq|k|-1/2\geq|k|/2$ and then $k^2\geq|k|$ for $k\neq0$. This estimate bounds the coherent sum of all tails, including their overlaps, and is independent of the input state.

\subsubsection{Evolution near zero energy offset}

The offset drift $U_x(\theta)=\exp[-\mathrm{i}\theta(N+x)^2]$ is already defined in Sec.~\hyperref[sec:finite]{IV}. Write the same fixed sequence on this ladder as
\begin{equation}
    \label{eq:fixed_offset_sequence}
    W_\ell(x)=U_x(\theta_\ell)A(g_\ell)\cdots U_x(\theta_1)A(g_1),\qquad W_\ell(0)=W_\ell.
\end{equation}
This is the sequence on the left-hand side of Eq.~(\ref{eq:offset_sequence}); the added argument only records the energy offset. For any fixed $\ket\psi=\sum_nc_n\ket n\in\mathcal H$,
\begin{equation}
    \label{eq:offset_drift_convergence}
    \left\|[U_x(\theta)-U(\theta)]\ket\psi\right\|^2=\sum_n|c_n|^2\left|\exp[-\mathrm{i}\theta(n+x)^2]-\exp(-\mathrm{i}\theta n^2)\right|^2\longrightarrow0
\end{equation}
as $x\to0$. Each phase difference tends to zero and has squared magnitude at most $4$. A finite sideband range therefore controls the sum: the remaining tail is bounded by $4\sum_{|n|>M}|c_n|^2$, which can be made arbitrarily small, while every term inside the range tends to zero.

The difference of two finite sequence products can be expanded into terms containing one drift difference $U_x(\theta_j)-U(\theta_j)$. The factors to its right act on a fixed ideal-evolved state, and the factors to its left preserve the norm. Equation~(\ref{eq:offset_drift_convergence}) therefore gives
\begin{equation}
    \label{eq:offset_sequence_convergence}
    \left\|[W_\ell(x)-W_\ell]\ket\psi\right\|\longrightarrow0,\qquad\ket\psi\in\mathcal H.
\end{equation}
To obtain one error bound for every normalized input in $\mathcal E$, choose an orthonormal basis $\{\ket{e_a}\}_{a=1}^{d}$, where $d=\dim\mathcal E$. Cauchy--Schwarz gives
\begin{equation}
    \label{eq:uniform_offset_convergence}
    \left\|[W_\ell(x)-W_\ell]P_{\mathcal E}\right\|^2\leq\sum_{a=1}^{d}\left\|[W_\ell(x)-W_\ell]\ket{e_a}\right\|^2\longrightarrow0.
\end{equation}
This finite-basis estimate is the step that makes the convergence apply uniformly to all normalized inputs in $\mathcal E$.

We next average this error over the Gaussian energy distribution. For any fixed $0<x_0<1/2$, split the integral into $|x|\leq x_0$ and the remaining offsets. Both sequence operators are unitary, so their difference has norm at most $2$. Since $f$ is normalized,
\begin{equation}
    \label{eq:gaussian_weighted_error_bound}
    \begin{split}
        &\hbar\omega_0\int_{-1/2}^{1/2}\mathrm dx\,|f(x\hbar\omega_0,\sigma_E)|^2\left\|[W_\ell(x)-W_\ell]P_{\mathcal E}\right\|^2\\
        &\quad\leq\sup_{|x|\leq x_0}\left\|[W_\ell(x)-W_\ell]P_{\mathcal E}\right\|^2+4\hbar\omega_0\int_{x_0<|x|<1/2}\mathrm dx\,|f(x\hbar\omega_0,\sigma_E)|^2\\
        &\quad\leq\sup_{|x|\leq x_0}\left\|[W_\ell(x)-W_\ell]P_{\mathcal E}\right\|^2+\frac{4\hbar\omega_0}{\sqrt{2\pi}\sigma_E}\exp\left[-\frac{x_0^2(\hbar\omega_0)^2}{2\sigma_E^2}\right].
    \end{split}
\end{equation}
First choose $x_0$ small enough to control the first term using Eq.~(\ref{eq:uniform_offset_convergence}). For this fixed $x_0$, the second term tends to zero as $\sigma_E\to0$. Therefore,
\begin{equation}
    \label{eq:zeta}
    \hbar\omega_0\int_{-1/2}^{1/2}\mathrm dx\,|f(x\hbar\omega_0,\sigma_E)|^2\left\|[W_\ell(x)-W_\ell]P_{\mathcal E}\right\|^2\longrightarrow0.
\end{equation}

\subsubsection{Combining the central term and Gaussian tails}

Using Eq.~(\ref{eq:Q_fibers}) for the input and the ideal output, the difference between the two finite-width constructions on one offset ladder is
\begin{equation}
    \label{eq:continuum_error_decomposition}
    \begin{split}
        &\sum_n\ket n\bra{E_0+(n+x)\hbar\omega_0}\left(W_\ell^{\mathrm{cont}}Q_{\sigma_E}-Q_{\sigma_E}W_\ell\right)\ket\psi\\
        &\quad=f(x\hbar\omega_0,\sigma_E)[W_\ell(x)-W_\ell]\ket\psi\\
        &\qquad+\sum_{k\neq0}f((x+k)\hbar\omega_0,\sigma_E)\left[W_\ell(x)(b^\dagger)^k-(b^\dagger)^kW_\ell\right]\ket\psi.
    \end{split}
\end{equation}
The first term is the central Gaussian multiplied by the evolution error. In the remaining sum, unitarity bounds each of the two tail contributions by Eq.~(\ref{eq:tail_bound}). After integrating over offsets, we obtain
\begin{equation}
    \label{eq:intertwining}
    \begin{split}
        &\left\|\left(W_\ell^{\mathrm{cont}}Q_{\sigma_E}-Q_{\sigma_E}W_\ell\right)P_{\mathcal E}\right\|\\
        &\quad\leq2\sum_{k\neq0}\left[\hbar\omega_0\int_{-1/2}^{1/2}\mathrm dx\,|f((x+k)\hbar\omega_0,\sigma_E)|^2\right]^{1/2}\\
        &\qquad+\left[\hbar\omega_0\int_{-1/2}^{1/2}\mathrm dx\,|f(x\hbar\omega_0,\sigma_E)|^2\left\|[W_\ell(x)-W_\ell]P_{\mathcal E}\right\|^2\right]^{1/2}\longrightarrow0.
    \end{split}
\end{equation}
The two contributions vanish by Eqs.~(\ref{eq:tail_bound}) and~(\ref{eq:zeta}). This proves the second limit in Eq.~(\ref{eq:corollary_key_limits}), with the same width-dependent error bound for every normalized input in $\mathcal E$.

\subsection{Combining the ideal control and finite-width errors}
\label{sec:width_combination}

We now choose the ideal sequence first and the width second. It suffices to prove the result for $0<\varepsilon<1$. By Theorem~1, choose a finite sequence satisfying
\begin{equation}
    \label{eq:ideal_corollary_tolerance}
    \left\|(W_\ell-T)P_{\mathcal E}\right\|<\frac{\sqrt\varepsilon}{4}.
\end{equation}
Keep this sequence fixed. For every normalized $\ket\psi\in\mathcal E$, the finite-width output error separates into
\begin{equation}
    \label{eq:unnormalized_error_decomposition}
    W_\ell^{\mathrm{cont}}Q_{\sigma_E}\ket\psi-Q_{\sigma_E}T\ket\psi=\left(W_\ell^{\mathrm{cont}}Q_{\sigma_E}-Q_{\sigma_E}W_\ell\right)\ket\psi+Q_{\sigma_E}(W_\ell-T)\ket\psi.
\end{equation}
The first term comes from the finite energy width; the second is the broadened ideal control error. Taking norms gives
\begin{equation}
    \label{eq:unnormalized_error}
    \begin{split}
        &\left\|W_\ell^{\mathrm{cont}}Q_{\sigma_E}\ket\psi-Q_{\sigma_E}T\ket\psi\right\|\\
        &\quad\leq\left\|\left(W_\ell^{\mathrm{cont}}Q_{\sigma_E}-Q_{\sigma_E}W_\ell\right)P_{\mathcal E}\right\|+\|Q_{\sigma_E}\|\frac{\sqrt\varepsilon}{4}.
    \end{split}
\end{equation}
The evolution term tends to zero by Sec.~\hyperref[sec:width_dynamics]{V.B}, and $\|Q_{\sigma_E}\|\to1$ by Sec.~\hyperref[sec:width_normalization]{V.A}. Both estimates are independent of the normalized input.

It remains to normalize the two outputs. Unitarity gives $\|W_\ell^{\mathrm{cont}}Q_{\sigma_E}\ket\psi\|=\|Q_{\sigma_E}\ket\psi\|$. The reverse triangle inequality therefore gives
\begin{equation}
    \label{eq:output_norm_difference}
    \left|\|Q_{\sigma_E}T\ket\psi\|-\|Q_{\sigma_E}\ket\psi\|\right|\leq\left\|W_\ell^{\mathrm{cont}}Q_{\sigma_E}\ket\psi-Q_{\sigma_E}T\ket\psi\right\|.
\end{equation}
Insert and subtract $Q_{\sigma_E}T\ket\psi/\|Q_{\sigma_E}\ket\psi\|$ when comparing the normalized states. Equation~(\ref{eq:output_norm_difference}) bounds the additional normalization error by the same output error. Using Eq.~(\ref{eq:gram_bound}) to bound the input norm from below yields
\begin{equation}
    \label{eq:normalized_error}
    \begin{split}
        &\left\|W_\ell^{\mathrm{cont}}\ket{\Psi_{\psi,\sigma_E}}-\ket{\Psi_{T\psi,\sigma_E}}\right\|\\
        &\quad\leq\frac{2\left\|W_\ell^{\mathrm{cont}}Q_{\sigma_E}\ket\psi-Q_{\sigma_E}T\ket\psi\right\|}{\|Q_{\sigma_E}\ket\psi\|}\\
        &\quad\leq\frac{2\left(\left\|\left(W_\ell^{\mathrm{cont}}Q_{\sigma_E}-Q_{\sigma_E}W_\ell\right)P_{\mathcal E}\right\|+\|Q_{\sigma_E}\|\sqrt\varepsilon/4\right)}{\sqrt{1-\|Q_{\sigma_E}^\dagger Q_{\sigma_E}-\mathds1\|}}.
    \end{split}
\end{equation}
For sufficiently small width, the denominator is positive. The rightmost bound tends to $\sqrt\varepsilon/2$, so there is a $\sigma_E^*>0$ such that
\begin{equation}
    \label{eq:finite_width_distance}
    \left\|W_\ell^{\mathrm{cont}}\ket{\Psi_{\psi,\sigma_E}}-\ket{\Psi_{T\psi,\sigma_E}}\right\|^2<\varepsilon,\qquad0<\sigma_E<\sigma_E^*.
\end{equation}
This holds for every normalized input in $\mathcal E$. For the two normalized output states, their overlap obeys
\begin{equation}
    2-2\operatorname{Re}\braket{\Psi_{T\psi,\sigma_E}|W_\ell^{\mathrm{cont}}|\Psi_{\psi,\sigma_E}}=\left\|W_\ell^{\mathrm{cont}}\ket{\Psi_{\psi,\sigma_E}}-\ket{\Psi_{T\psi,\sigma_E}}\right\|^2<\varepsilon.
\end{equation}
Since $0<\varepsilon<1$, it follows that
\begin{equation}
    \left|\braket{\Psi_{T\psi,\sigma_E}|W_\ell^{\mathrm{cont}}|\Psi_{\psi,\sigma_E}}\right|^2>(1-\varepsilon/2)^2>1-\varepsilon.
\end{equation}
The sequence is fixed before the width is chosen, and the same sequence works for every normalized input in $\mathcal E$ and every $0<\sigma_E<\sigma_E^*$. The threshold may depend on $T$, $\mathcal E$, the selected sequence, and the requested accuracy. This proves Corollary~1.

{\section{Numerical methods and data}}\label{sec:numerical}

{For the ideal-ladder examples in Figs.~2--4, we optimize the $3\ell$ real parameters given by the magnitudes and phases of $g_j$ and the drift phases $\theta_j$, minimizing one minus the corresponding fidelity defined in the main text. We use the L-BFGS-B algorithm implemented in SciPy with analytic gradients obtained by backward propagation through the sequence. Multiple starting points are drawn from scrambled Sobol sequences; promising solutions are further refined by local optimization and perturbations of their parameters. The coupling bounds are $|g_j|\leq2$ for Figs.~2 and 3 and $|g_j|\leq1$ for Fig.~4. For the ideal ladder, the coupling and drift phases can be represented in $[-\pi,\pi)$ and $[0,2\pi)$, respectively, by periodicity. The reported fidelities are those of the best solutions found, without a guarantee of global optimality.}

{For the finite-width examples in Fig.~5(a)--(b), the objective is $1-\mathcal F_{\sigma_E}$, with the sequence optimized separately for each incident energy width and stage count. We use the same L-BFGS-B algorithm with analytic gradients and scrambled Sobol initial-point selection described above for the ideal-ladder examples. The upper bound on each coupling magnitude $|g_j|$ is initially set to $2.5$. If an optimized coupling magnitude reaches this bound, the bound is increased to $3$ and the optimization is repeated. The upper bound on each drift phase $\theta_j$ is initially set to $0.1\pi$. Whenever an optimized drift phase reaches the current upper bound, this bound is increased by $0.1\pi$ and the optimization is repeated. In contrast to the ideal ladder, the drift phases in this calculation cannot be reduced modulo $2\pi$.}

The accompanying file \texttt{PINEMAD\_parameters.csv} contains the PINEMAD sequence parameters and fidelities for Figs.~2--4 and 5(a)--(b), together with the fixed-parameter reference results in Table~\ref{tab:fixed_parameters}. Each row specifies one stage $j$, with $j=1$ acting first:
\begin{equation}
    W_\ell =U(\theta_\ell)A(g_\ell)\cdots U(\theta_1)A(g_1).
\end{equation}

The columns \texttt{figure}, \texttt{target}, \texttt{ell}, and \texttt{Delta\_E\_FWHM\_eV} jointly identify a calculation. The value \texttt{main-text table} in the \texttt{figure} column is a retained label for the fixed-parameter reference results now presented in Table~\ref{tab:fixed_parameters}. A zero in \texttt{Delta\_E\_FWHM\_eV} denotes an ideal-ladder calculation, while a nonzero value specifies the FWHM $\Delta E$ of the incident energy probability distribution, in eV. The column \texttt{stage\_j} gives the stage index. Columns \texttt{g\_real} and \texttt{g\_imag} give $\operatorname{Re}g_j$ and $\operatorname{Im}g_j$, and \texttt{theta\_rad} gives $\theta_j$ in radians. The column \texttt{total\_theta\_rad} gives the total drift phase $\Theta=\sum_{j=1}^{\ell}\theta_j$ in radians and repeats the same value for every stage of a sequence. The achieved fidelity is recorded in \texttt{fidelity} and repeated for every stage of the same sequence. Parameters and fidelities are supplied at their stored precision.

For comparison, Table~\ref{tab:fixed_parameters} lists the finite-width fidelities obtained by applying the ideal-ladder sequences of Fig.~2 without reoptimizing their parameters.

\begin{table}[htbp]
\centering
\caption{State-preparation fidelities obtained by applying the ideal-ladder PINEMAD parameters of Fig.~2 to finite-width incident wavepackets without reoptimization. The optical wavelength is $1030\,\mathrm{nm}$, and $\Delta E$ is the FWHM of the incident Gaussian energy probability distribution.}
\label{tab:fixed_parameters}
\begin{ruledtabular}
\begin{tabular}{ccc}
$\Delta E$ (eV) & $\mathcal F_{\sigma_E}$ (Target I) & $\mathcal F_{\sigma_E}$ (Target II) \\
0.1 & $6.29979\times10^{-2}$ & $1.76908\times10^{-3}$ \\
0.2 & $9.20029\times10^{-3}$ & $2.07312\times10^{-4}$ \\
0.3 & $1.65971\times10^{-3}$ & $2.93341\times10^{-4}$ \\
0.4 & $1.98970\times10^{-4}$ & $3.39008\times10^{-4}$ \\
0.5 & $2.93484\times10^{-5}$ & $3.01457\times10^{-4}$
\end{tabular}
\end{ruledtabular}
\end{table}

All finite-width calculations use a $1030\,\mathrm{nm}$ optical wavelength and an unchirped incident envelope as shown in the main text. To reproduce the full-wavepacket fidelities by direct propagation, write $x=n+\delta$, retain $n=-112,\ldots,112$, and sample $\delta\in[-1/2,1/2)$ at 1024 uniformly spaced midpoints. At each $\delta$, apply Eq.~(\ref{eq:bessel}) with shifts from $-36$ to $36$, followed by $\exp[-\mathrm{i}\theta_j(n+\delta)^2]$. Construct the target as a coherent sum of translated Gaussians with the coefficients specified in the main text, normalize it, and take the overlap in Eq.~(\ref{eq:full_fidelity}). Discrete Target II coefficients below magnitude $10^{-14}$ may be omitted at the reported precision. Using $|n|\leq160$ and Bessel shifts $|q|\leq48$ reproduces the ideal-example fidelities.

For a specified central electron kinetic energy $E_0$, the total physical drift length follows from Eq.~(\ref{eq:drift_length}) as $\sum_jL_j=\Theta L_T/(2\pi)$. Figure~5(c) uses the sequence for Target I with $\Delta E=0.4\,\mathrm{eV}$ and $\ell=9$. The actual drift phases stored in the file must be used without reduction modulo $2\pi$.

\bibliography{main}